\documentclass[
 reprint,
superscriptaddress,
groupedaddress,
 amsmath,
 amssymb,
 aps,
]{revtex4-2}

\usepackage{graphicx}% Include figure files
\usepackage{dcolumn}% Align table columns on decimal point
\usepackage{bm}% bold math
\usepackage{caption}
\usepackage{tikz}
\usepackage{tikz-network}
\usepackage{relsize}
\usepackage{tabularray}
\usepackage[figuresleft]{rotating}
\usepackage{lipsum}
\usepackage{cleveref}
\usepackage{float}
\restylefloat{table}
\usepackage{subfig}

\begin{document}

\preprint{APS/123-QED}

\title{Asymmetric Interactions Shape Survival During Population Range Expansions}

\author{Jason M. Gray$^{1,2}$, Rowan J. Barker-Clarke$^{2}$, Jacob G. Scott$^{1,2,3}$, Michael Hinczewski$^{1}$\\
$^1$Department of Physics, Case Western Reserve University, Cleveland, OH, 44106, USA\\
$^2$Translational Hematology \& Oncology Research, Cleveland Clinic, Cleveland OH, 44106, USA\\
$^3$Case Western Reserve University School of Medicine, Cleveland, OH, 44106, USA
}

% \author{Jason M. Gray}
% \affiliation{Translational Hematology \& Oncology Research, Cleveland Clinic, Cleveland OH, 44106, USA}
% \affiliation{Department of Physics, Case Western Reserve University, Cleveland, OH, 44106, USA}

% \author{Rowan J. Barker-Clarke}
% \affiliation{Translational Hematology \& Oncology Research, Cleveland Clinic, Cleveland OH, 44106, USA}

% \author{Michael Hinczewski}
% \affiliation{Department of Physics, Case Western Reserve University, Cleveland, OH, 44106, USA}
% \affiliation{michael.hinczewski@case.edu}

% \author{Jacob G. Scott}
% \affiliation{Translational Hematology \& Oncology Research, Cleveland Clinic, Cleveland OH, 44106, USA}
% \affiliation{Department of Physics, Case Western Reserve University, Cleveland, OH, 44106, USA}
% \affiliation{Case Western Reserve University School of Medicine, Cleveland, OH, 44106, USA}
% \affiliation{scottj10@ccf.org}

\date{\today}

\begin{abstract}
An organism that is newly introduced into an existing population has a survival probability that is dependent on both the population density of its environment and the competition it experiences with the members of that population. 
Expanding populations naturally form regions of high and low density, and simultaneously experience ecological interactions both internally and at the boundary of their range. 
For this reason, systems of expanding populations are ideal for studying the combination of density and ecological effects.
Conservation ecologists have been studying the ability of an invasive species to establish for some time, attributing success to both ecological and spatial factors.
Similar behaviors have been observed in spatially structured cell populations, such as those found in cancerous tumors and bacterial biofilms.
In these scenarios, novel organisms may be the introduction of a new mutation or bacterial species with some form of drug resistance, leading to the possibility of treatment failure.
In order to gain insight into the relationship between population density and ecological interactions, we study an expanding population of interacting wild-type cells and mutant cells. 
We simulate these interactions in time and study the spatially dependent probability for a mutant to survive or to take over the front of the population wave (gene surfing). 
Additionally, we develop a mathematical model that describes this survival probability and find agreement when the payoff for the mutant is positive (corresponding to cooperation, exploitation, or commensalism).
By knowing the types of interactions, our model provides insight into the spatial distribution of survival probability.
Conversely, given a spatial distribution of survival probabilities, our model provides insight into the types of interactions that were involved to generate it. 
\end{abstract}
\flushbottom
\maketitle
\section{Introduction}
\begin{figure*}[ht]
    \centering
    \includegraphics[width=\textwidth]{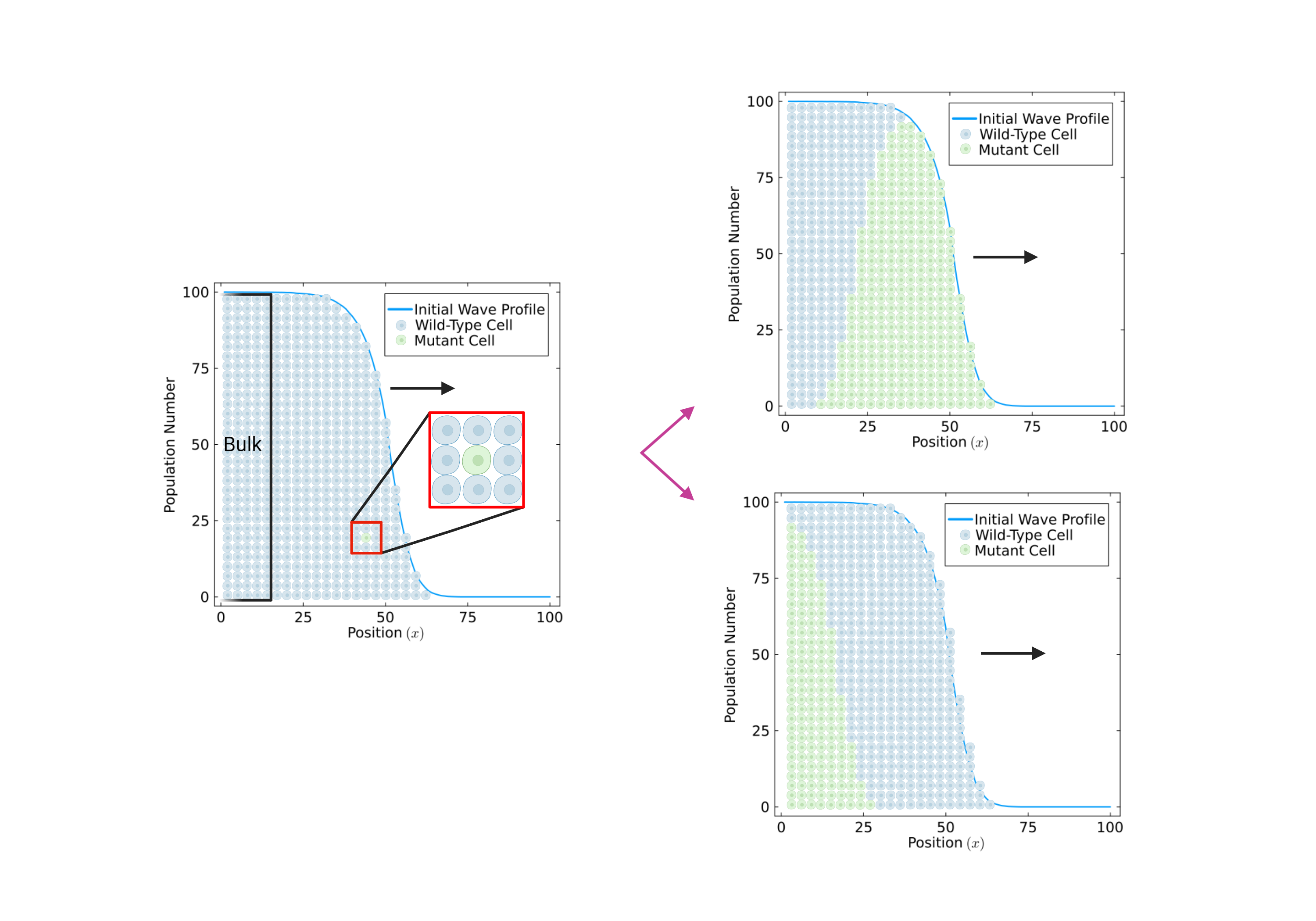}
    \captionsetup{justification=raggedright}
    \caption{\textbf{A mutant may survive by either surfing or abiding in an expanding population}. (Left) A mutant is initiated in an expanding population of wild-types. 
    The bulk of the wave corresponds to a uniform region of cells at carrying capacity (in this case it is $C=100$). 
    (Right) The wild-types and mutants interact, and both cell types interact with vacancies.
    After infinite time, there are two possibilities for survival shown in the right-hand side of the figure.
    (Top right) The mutant surfs, taking over the front of the population expansion.
    (Bottom Right) The mutant
    abides, surviving within the bulk of the expanding population, but is outpaced by the wild-types.}
    \label{fig:RE-surf-cartoon}
\end{figure*}

The spatial distribution of organisms is inextricable from population dynamics in most scenarios. In many cases, the assumption of a non-spatial, well-mixed system gives drastically different outcomes in terms of fixation probabilities, fixation times, and the rate of evolution \cite{frean2013effect,ohtsuki2006replicator,marrec2021toward,yagoobi2021fixation,kaznatcheev2015edge,ottino2017takeover,sharma2022suppressors}.
When there is high motility and mixing however, the approximation is most reasonable.
In the absence of these properties, there is a trade-off for analytical tractability.
The spatial structures of populations are dynamic in general.
One of the best examples of the dynamical nature of spatial population structure is a population expanding its range.
Range expansion at the most fundamental level is the result of two effects: demography and dispersal.
Demography has to do with the division of cells or the birth of offspring and the composition of a population.
Dispersal has to do with spatial migration, which can be short- or long-range, fast or slow in speed.
Together, these effects result in a spatially modulated population wave, with a higher density towards the bulk of the wave and a lower density towards the moving front. 
The simplest system with of a single type of asexually reproducing organism is described by a Fisher wave with average $\langle w \rangle(x)$ and a minimum wave speed $s > \sqrt{rD}$, where $r$ is an intrinsic growth rate and $D$ is a diffusion constant \cite{ablowitz1979explicit}.
Mathematical theory has been developed to describe the speed of expanding populations with different demographic and dispersal traits \cite{skellam1951random, shigesada1997biological, kot1996dispersal, hallatschek2009fisher, brunet2001effect, hallatschek2014acceleration}.
The density modulation has interesting consequences for the spatial and temporal heterogeneity of organisms in expanding populations as well.
One of the most notable features of range expansions is the observed enhancement of genetic drift at the wave-front resulting in (possibly successive) takeovers at the front: the phenomenon called ``gene surfing'' \cite{hallatschek2008gene, excoffier2008surfing, edmonds2004mutations, fusco2016excess, paulose2020impact,excoffier2009genetic,lehe2012rate}. 
Considering a single mutation arising in a wild-type wave, there are two situations that can occur: either a mutant is able to have enough progeny where the mutant population survives or all of the mutants die out.
If the mutant population has survived, as is depicted in Fig (\ref{fig:RE-surf-cartoon}), then it has either taken over the front of the expanding population (surfing) or it has established a stable population within the bulk of the wave with the possibility of being left behind by the front.
For brevity, we will call the latter situation ``abiding''.
What has been observed in theoretical models and experiments in many cases is the development of ``sectors'', where homogeneous regions of the same cell type (based on fluorescent labeling of neutral or non-neutral mutations) develop as the population expands \cite{korolev2010genetic,hallatschek2007genetic,weinstein2017genetic,aif2022evolutionary,kayser2019collective,golding1999studies}. 
Ecological interactions are an additional, but not necessarily orthogonal, component to determining survival probabilities, heterogeneity, and speed of range expansions.
Interest in the inclusion of ecological effects during range expansions has been increasing across many fields in biology.
These include, but are not limited to, conservation ecology, cancer biology, and biofilms research.
In light of the effects of anthropogenic climate change on habitat modification, conservation biologists are seeing increasing population migrations and changes to the types of ecological interactions experienced \cite{wallingford2020adjusting,pecl2017biodiversity,filbee2005ecological,erazo2024contribution}.
The usefulness of certain traits or strategies for a given invader is often assumed rather than quantified, but some progress has been made in this area \cite{estrada2016usefulness, maclean2017species}.
Although one may think of ecological interactions at the level of large-scale ecosystems, they also exist at the level of cells.
Some theoretical work has been done on ecological interactions of expanding asexual populations, focusing mainly on aspects of cooperation versus competition \cite{korolev2011competition, lavrentovich2014asymmetric,kayser2018emergence,le2003adaptive,narayanan2024mutualisms,nadell2016spatial}. 
In the context of cooperation versus competition, competition between different cell types excludes neighbors of those different types (competitive exclusion), thus creating regions of homogeneous cells. 
Microorganisms such as bacteria often live in complex spatial structures called biofilms.
Biofilms can be attached to a surface (e.g. staphylococci, which are recognized as the most frequent cause of biofilm associated infections \cite{silva2021biofilm, mader2021building, lebeaux2013vitro}). 
The extracellular matrix in biofilms has been found to confer some level of protection against antibiotics, making the development of resistance mutations more likely \cite{kwon2008higher}.
Surface-attached biofilms are commonly seen in implanted prosthetic devices such as catheters and often cause infections as a result \cite{bouhrour2024medical,harding2014combating}.
The cells within biofilms interact and influence each others fitness through social phenotypes \cite{nadell2008sociobiology}. 
Social phenotypes are often cooperative in nature, for example: secretion of nutrient chelators (molecules that surround a nutrient to prevent it from unwanted reaction processes and help them become more soluble) \cite{visca2007pyoverdine,griffin2004cooperation},
digestive enzymes (which reduce energy expenditure requirements, increasing metabolic efficiency) \cite{allison2005cheaters}, surface adhesins (which help bacteria attach to a surface, and reduce the effects of damaging shear forces) \cite{absalon2011communal,xavier2011molecular}, wetting agents (which help to deliver bacteria to hydrophobic surfaces) \cite{matsuyama1996bacterial,werb2017surface}, structural polymers (which help in the formation of the biofilm itself) \cite{flemming2010biofilm}, and signaling molecules (such as quorum sensing molecules used to coordinate gene expression) \cite{west2007social}.
Social phenotypes may also be antagonistic or competitive \cite{hibbing2010bacterial,rendueles2015mechanisms,burmolle2014interactions}, for example: by hoarding nutrients \cite{vance1984interference}, blocking access to favorable habitats (such as binding sites on a surface, or regions with high nutrient density) \cite{case1974interference}, forcing the dispersal of competitors (thus reducing cooperative benefits) \cite{nadell2008evolution}, and producing antimicrobial toxins (e.g. in T6SS dueling of P. aeruginosa 
) \cite{basler2013tit,ho2014view}.
Additionally, spatial gradients of nutrients, salinity, pH, antibiotics, or chemotherapy modify the environment and may trigger expansion events and invasion. 
In these environments, mutations may arise that adapt to harsher conditions. 
This is an issue in treatments of cancer, viral and bacterial infections, and autoimmune conditions, because a mutation may arise that is resistant to a drug therapy (e.g., antibiotics or chemotherapies).
These types of mutations may be more sensitive to genetic drift early on, since the strength of stochastic fluctuations is larger than or comparable to the number of mutants.
If ecological interactions are able to increase the fitness of the mutant from its intrinsic fitness, resistance mutations may be able to overcome the initial highly stochastic phase, allowing it to establish in the population \cite{kaznatcheev2015edge, kaznatcheev2019fibroblasts, farrokhian2022measuring, sharma2021spatial, maltas2024frequency, barker2023balance}.
Here, we seek to combine the effects of ecological interactions with the effects of density dependence in order to develop a more complete picture of survival during range expansions. 
We explore the phase space of asymmetric game interactions in order to see their effects on the survival of beneficial, neutral, and deleterious mutants in the context of range expansions.
We compare the results of stochastic simulations to a mathematical model for the survival probability and find great agreement when the survival probability of a mutant deep in the bulk is greater than the survival probability far ahead of the wave-front.
\section{Asymmetric Games in a Reaction-Diffusion System}
A common way to model interactions is through the use of game theory.
Game theory has been shown to be a useful tool in the description of interactions in biological systems, economics, and the social sciences.
At the fundamental level, it includes ``players'' and ``strategies''.
Although these terms are rather anthropomorphic, they simply correspond to individual agents and their available set of actions that they can perform.
An individual player may play a strategy against another, which results in a payoff for the players.
One of the simplest types of games is a symmetric game, represented of these by a ``payoff matrix'':
\begin{equation}
    P=\begin{pmatrix}
        P_{AA} & P_{AB} \\
        P_{BA} & P_{BA}
    \end{pmatrix}.
\end{equation}
Here, our players can choose to play strategy $A$ or strategy $B$.
In general, if a player plays a strategy ($I$) against the other player's strategy ($J$), they receive the payoff $P_{IJ}$. 
Game theory applied to evolutionary systems (Evolutionary Game Theory) is interested in the evolution of the distribution of strategies over time \cite{smith1973logic}. 
Additionally, the identity of a player is the strategy that they play in EGT.
For example, in a population of a mix of $A$-strategists and $B$-strategists, one would be interested in how the amounts of each strategy change over time.
One of the most common ways to model this system is through the replicator equation:
\begin{equation}
    \frac{dx_i}{dt} = x_i\Bigg(\mathlarger{\sum}\limits_{j=1}^S P_{ij}x_j-\mathlarger{\sum}\limits_{j=1}^S\mathlarger{\sum}\limits_{k=1}^S x_jP_{jk}x_k\Bigg),
\end{equation}
where $x_i$ represents the proportion of strategy $i$, $\vec{x}$ represents the vector of proportions of every strategy, and $P$ represents the payoff matrix. 
It is important to note here that the payoff has units of inverse time $[t]^{-1}$ and may be interpreted as a rate.
The sum $f_i=\sum_{j=1}^S P_{ij}x_j$ is the frequency dependent fitness of the strategy $i$ and the double sum $\phi=\sum_{j=1}^S\sum_{k=1}^S x_jP_{jk}x_k$ is the average fitness of the population.
The original replicator equation is derived from a modified exponential growth model:
\begin{equation}
    \frac{dN_i}{dt} = N_i\mathlarger{\sum}\limits_{j=1}^S P_{ij}x_j,
\end{equation}
where $N_i$ represents the number of strategy $i$, and $x_i=N_i/N_{tot}$. 
When $N_{tot}$ is a fixed carrying capacity, the replicator equation is without a second term corresponding to the average fitness of the population.
The fixed carrying capacity replicator equation is then:
\begin{equation}\label{eq:fcc_rep}
    \frac{dx_i}{dt} = x_i\mathlarger{\sum}\limits_{j=1}^S P_{ij}x_j
\end{equation}
In the replicator equation above, one may notice that there is no notion of variability in payoff from any given interaction coming from individual differences.
In nature however, differences in phenotype (size, strength, speed, intellect, etc.) or life history (injuries, energy spent, arrival time, etc.) among other things may create an asymmetry in payoffs.
The representation of an asymmetric game is given by a ``payoff bimatrix'' or ``payoff tensor''.
The case of a two-player asymmetric game, where each player may play one of two strategies $A$ and $B$, gives the following payoff tensor:
\begin{equation}
P=\begin{pmatrix}
    (P_{AA}^{1},P_{AA}^{2}) & (P_{AB}^{1},P_{AB}^{2})  \\
    (P_{BA}^{1},P_{BA}^{2}) & (P_{BB}^{1},P_{BB}^{2})
\end{pmatrix}.
\end{equation}
Here, the strategies are generically labeled as ``A'' and ``B'' and the players are labeled as $1$ and $2$. 
To see how payoffs are attributed, we will look at the component $P_{AB}^1$. 
Starting with the subscripts, we see that the first player played strategy $A$ and the second player played strategy $B$. The superscript $1$, means that the first player gets the payoff of $P_{AB}^1$ for this interaction. 
Player two would then receive the payoff $P_{AB}^2$ for the same interaction. 
Conversely, if player one now plays strategy $B$ and player two plays strategy $A$, then player one receives the payoff $P_{BA}^1$ and player two receives the payoff $P_{BA}^2$.
In this context, setting $P_{IJ}^1=P_{JI}^2$ for all sets of $I$ and $J$, gives us a symmetric game. 
For example, if we had $P_{AB}^1=P_{BA}^2$, this means that if any player plays strategy $A$ (player $1$ on the left side and player $2$ on the right side) and the other player plays strategy $B$ (player $2$ on the left side and player $1$ on the right side), they will get the same payoff value, regardless of their identity.
In EGT, one often uses a wild-type and various mutations as the strategies/identites in the system, and they are interested in how the numbers of the wild-type and various mutants changes over time.
For a two-player asymmetric game, we may then write the payoff tensor as:
\begin{equation}
P=\begin{pmatrix}
    (P_{WW}^{W},P_{WW}^{W}) & (P_{WM}^{W},P_{WM}^{M})  \\
    (P_{MW}^{M},P_{MW}^{W}) & (P_{MM}^{M},P_{MM}^{M})
\end{pmatrix}.
\end{equation}

In order to make sense of the distinction between different orderings of the subscripts, we interpret them as a temporal ordering.
For example, the components $P_{MW}^{M}$ and $P_{MW}^{W})$ would correspond to a mutant initiating the interaction and the wild-type accepting the interaction. 
The mutant initiates the interaction by releasing a ligand or metabolite, changing its fitness (e.g. by expending energy to produce the molecule).
The molecule is then bound to a receptor on the wild-type or is taken into its interior, changing its fitness (e.g. by stimulating growth). 
In contrast, the components $P_{WM}^{W}$ and $P_{WM}^{M})$ would correspond to a wild-type initiating the interaction and a mutant accepting the interaction. 
In biological contexts, these types of interactions are split into: ``autocrine'' interactions, where a released molecule interacts with the same cell that released it, ``paracrine'' interactions, where a released molecule interacts with a different cell than the cell that released it without direct contact, and ``gap junction'' interactions, where a molecule interacts with a different cell than released via a gap junction contact.
If there is no difference between cells of the same type (wild-type or mutant), then the game is only asymmetric if $P_{MW}^M \ne P_{WM}^{W}$ and/or $P_{WM}^M \ne P_{MW}^{W}$. 
This means that there is a fundamental difference between mutants and wild-types in terms of how they interact with each other.
Rather than writing chemical reactions for every possible intermediary molecule that can be transferred between a wild-type and a mutant, we write a reduced model of reactions between the cells themselves corresponding to the net effect. 
According to the replicator equation, the payoff tensor elements have units of inverse time $[t]^{-1}$, we therefore take the magnitude of the payoff tensor elements to be the rates of the cell reactions.
Additionally, we consider a time scale in which cell numbers only go up by one (division) or down by one (cell death).
In order for a cell to divide, there must be space for it to divide.
For populations with a finite carrying capacity ($C$), we then have $P_{IJ}^K=-P_{JI}^K$.
In order to encode this in the form of a reaction, we use the sign of the payoff tensor component to determine which cell divides and which cell dies. 
We then write the reactions where a wild-type sends signals to a mutant in the following way:
\begin{align}\label{eq:reac1}
    \mathcal{W} + \mathcal{M} \xrightarrow{|P_{WM}^M|} (1+\textrm{sgn}(P_{WM}^W))\mathcal{W} \\+ (1+\textrm{sgn}(P_{WM}^M))\mathcal{M}\nonumber,
\end{align}
where $\mathcal{W}$ represents a wild-type cell, $\mathcal{M}$ represents a mutant cell. 
Here, $\textrm{sgn}(x)=1$ if $x>0$, $\textrm{sgn}(x)=-1$ if $x<0$, and $\textrm{sgn}(x)=0$ if $x=0$.
The stoichiometric components are then given by the values of $\textrm{sgn}(x)$.
If the interaction is beneficial for the mutant, we have $P_{WM}^M > 0$, $P_{WM}^W < 0$, and the resulting reaction:
\begin{equation}\label{eq:reac2}
    \mathcal{W} + \mathcal{M} \xrightarrow{|P_{WM}^M|} 2\mathcal{M} ,
\end{equation}
where the wild-type has died and the mutant has divided.
Conversely, if the interaction is detrimental for the mutant, we have $P_{WM}^M < 0$, $P_{WM}^W > 0$, and the resulting reaction:
\begin{equation}\label{eq:reac3}
    \mathcal{W} + \mathcal{M} \xrightarrow{|P_{WM}^M|} 2\mathcal{W},
\end{equation}
where the mutant has died and the wild-type has divided.
We may write a set of ODEs corresponding to the mass action law using the rates and stoichiometric components for this simplified system:
\begin{align}
    \frac{dw}{dt} &= \textrm{sgn}(P_{WM}^W)|P_{WM}^W|wm\nonumber\\
                    &= P_{WM}^W wm
\end{align}
and 
\begin{align}
    \frac{dm}{dt} &= \textrm{sgn}(P_{WM}^M)|P_{WM}^M|wm\nonumber\\
                    &= P_{WM}^M wm,
\end{align}
where $w$ represents the fraction of wild-types and $m$ represents the fraction of mutants.
Note that these equations match the form of the fixed carrying capacity replicator equation (Eq. \ref{eq:fcc_rep}).

In our model, we are interested in the expansion of a population in space.
We implement spatial components in our model in two ways: the addition of vacancies ($V$) which represent empty space, and the possibility of swapping between neighboring locations.
Interactions take place locally (at the same location $i$) and swapping takes place non-locally (between  $i$ and $i\pm 1$).
The local payoff tensor is given by:
\begin{equation}
P=\begin{pmatrix}
    (0,0,0) & (P_{WM}^{W},P_{WM}^{M},0) & (P_{WV}^{W},0,P_{WV}^{V})  \\
    (P_{MW}^{W},P_{MW}^{M},0) & (0,0,0) & (0,P_{MV}^{M},P_{MV}^{V})  \\
    (P_{VW}^{W},0,P_{VW}^{V}) & (0,P_{VM}^{M},P_{VM}^{V}) & (0,0,0) 
\end{pmatrix}.
\end{equation}
Here, we set the self-interaction terms to zero, as our focus is on the wild-type and mutant interactions.
Through interaction with vacancies, we can encode intrinsic division and death events which in combination give an intrinsic growth rate.
We encode the birth events into the $P_{IV}$ components of the local payoff tensor by setting $P_{IV}^I = 1$.
We then encode the death events into the $P_{VI}$ components with $P_{VW}^W = -(1-r_w)$ and $P_{VM}^M = -(1-r_m)$, where $r_w$ and $r_m$ are the intrinsic growth rates of the wild-type and mutant respectively.
By writing the birth and death payoff components in this way, we only need to modify two parameters to compare the effects of modifying intrinsic growth rates with a fixed game interaction.
Writing out the mass action ODE that describe the dynamics of the mutant at a single location we would have:
\begin{align}
\frac{dm_i}{dt} &= (P_{VM}^M+P_{MV}^M)v_im_i\\
                &+(P_{WM}^M+P_{MW}^M)w_im_i \nonumber\\
                    &= (-(1-r_m)+1)v_im_i \nonumber \\
                    &+(P_{WM}^M+P_{MW}^M)w_im_i \nonumber \\ 
                    &= r_m v_im_i+(P_{WM}^M+P_{MW}^M)w_im_i \nonumber,
\end{align}
where $v_i$, $m_i$, and $w_i$ are the fraction of vacancies, mutants, and wild-types at location $i$ respectively.
We may define a swapping tensor ($S$) and write similar reaction equations in order to incorporate the effects of diffusion (see Eqs. A$5-$A$15$ in the appendix). 
Moving from a discrete to a continuous space ($\Delta x \rightarrow 0$), the resulting reaction-diffusion equation for mutants at location $x$ is given by:
\begin{equation}\label{eq:det_ode}
\frac{dm}{dt} = r_mvm+(P_{WM}^M+P_{MW}^M)wm + \frac{d^2 m}{dx^2}, 
\end{equation}
where we have set the diffusion constant\\ $D=1[d]^2[t]^{-1}$, where $[d]$ are units of distance.
We will use this equation in the derivation of an ODE for the survival probability of a new mutation appearing at location $x$.

\begin{table*}
    \centering
    \caption{Possible events in our model.}
    \label{tab:events}
    \begin{tblr}{|l|c| } 
        \hline
        \textit{Event Description} & \textit{Illustration}\\ 
        \hline
        A wild-type is born at location $i$. & \includegraphics[scale=0.25]{"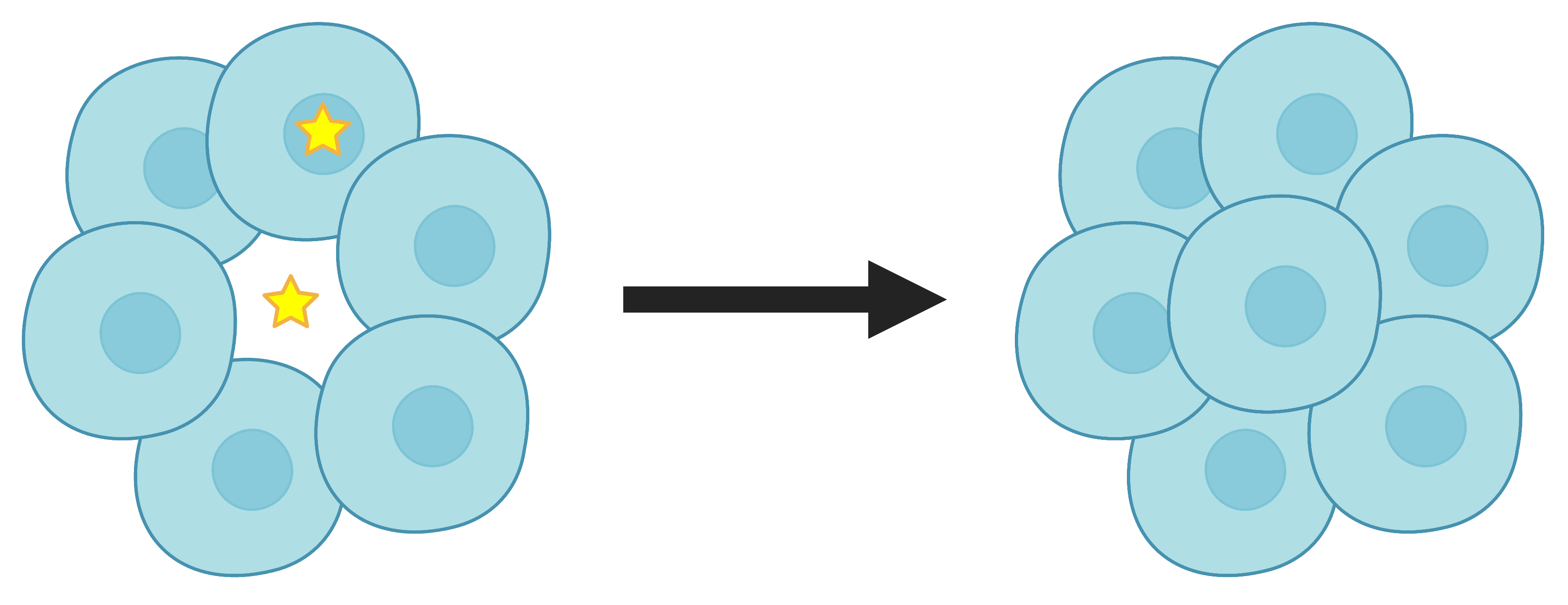"}\\
        \hline[dashed]
       A wild-type dies at location $i$. & \includegraphics[scale=0.25]{"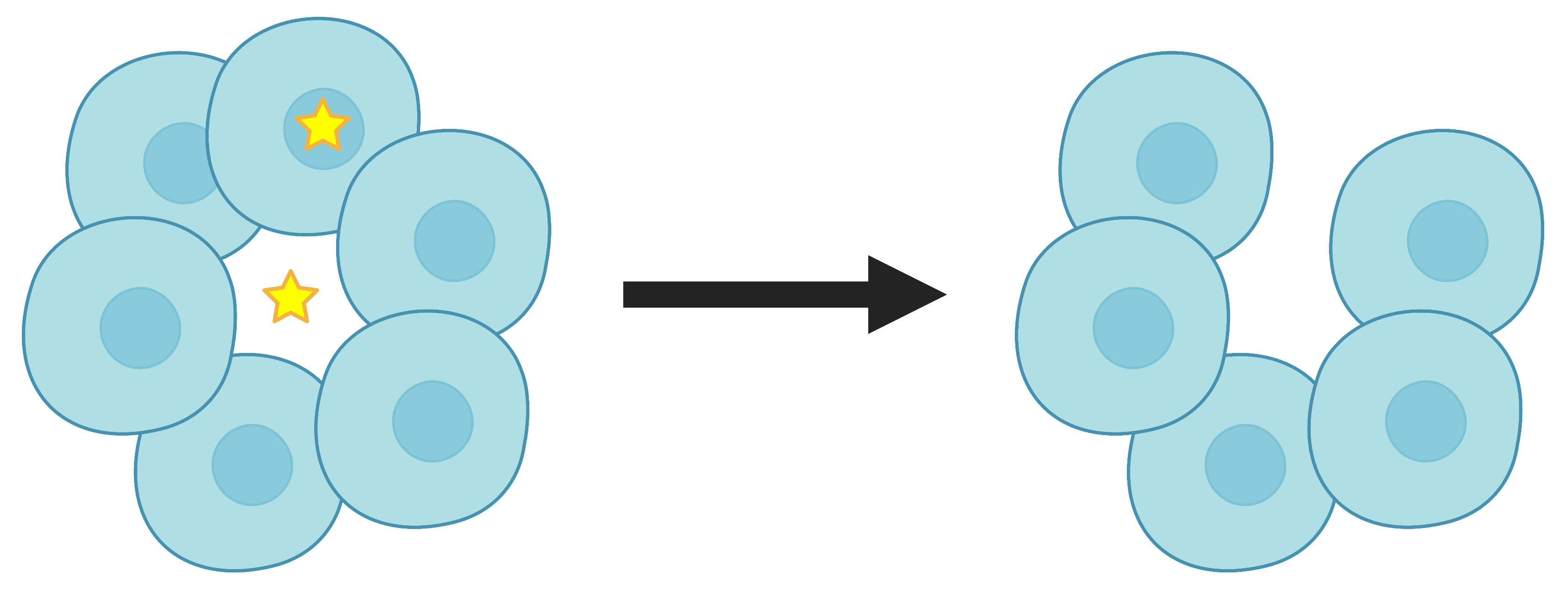"}\\
        \hline[dashed]
        A mutant is born at location $i$. & \includegraphics[scale=0.25]{"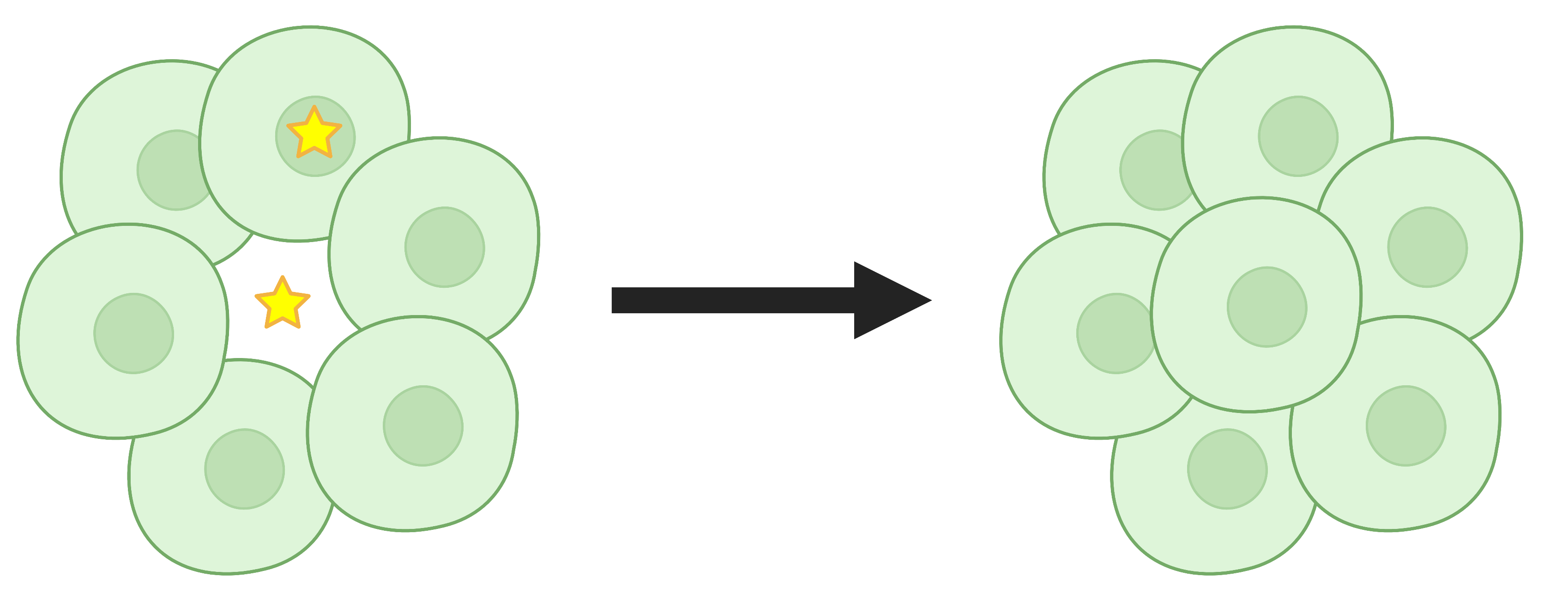"}\\
        \hline[dashed]
       A mutant dies at location $i$. & \includegraphics[scale=0.25]{"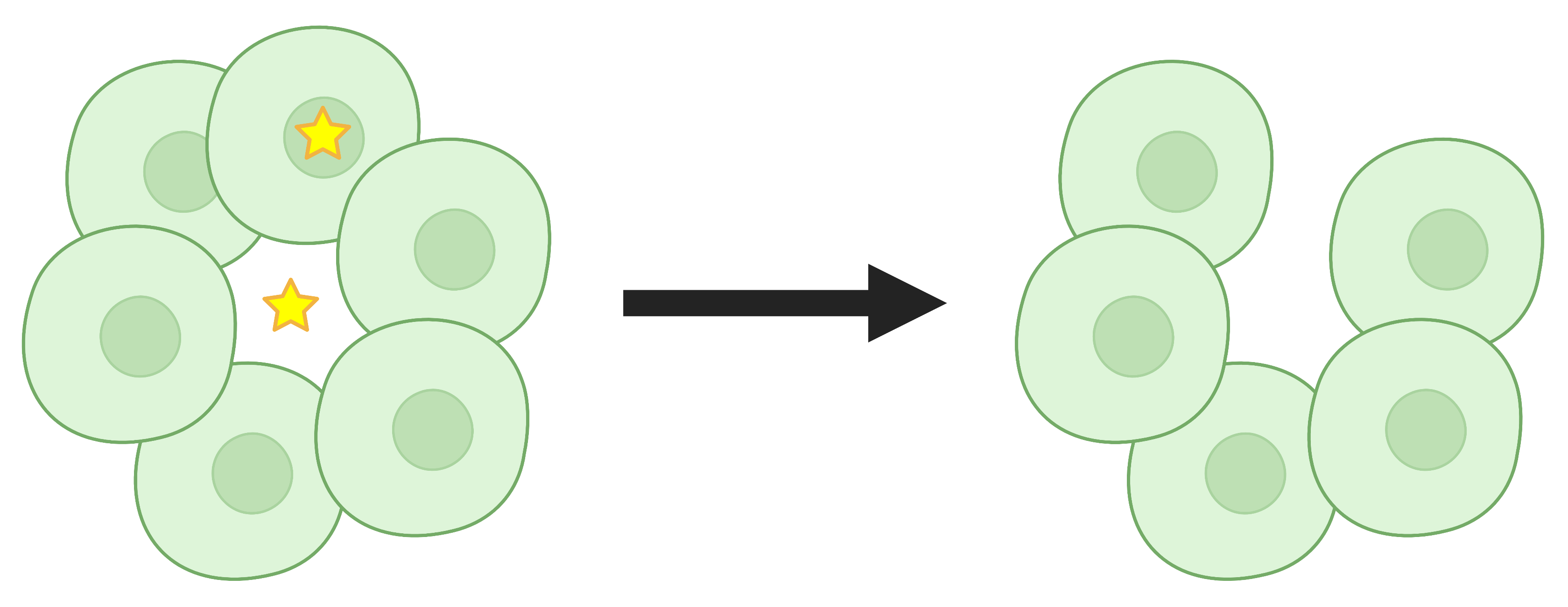"}\\
        \hline[dashed]
        A wild-type initiates an interaction with \newline
        a mutant at location $i$. & \SetCell[r=2]{m,7cm}\includegraphics[scale=0.25]{"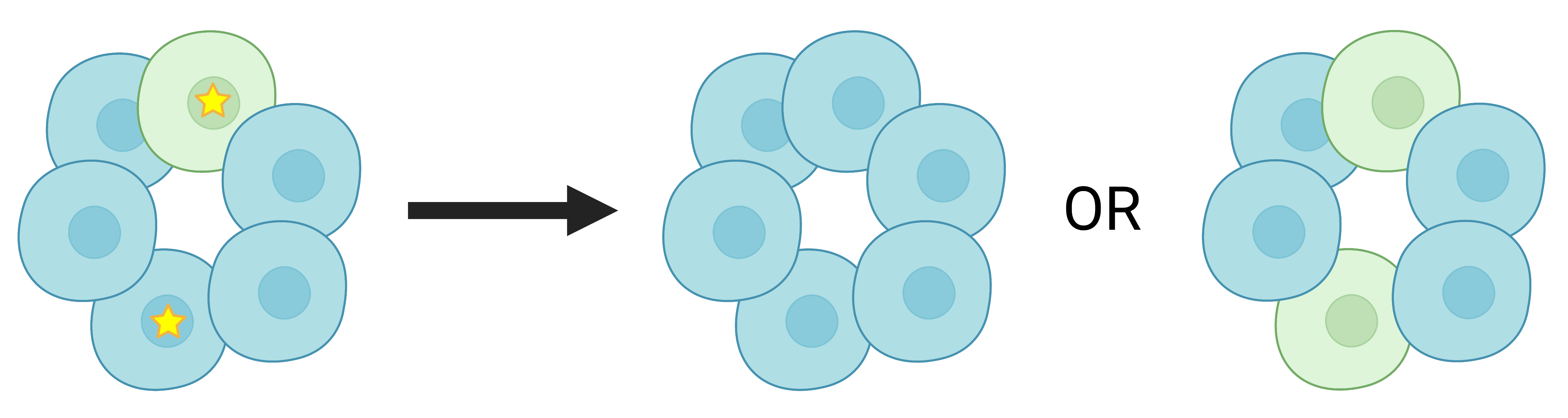"}\\
        \hline[dashed]
        A mutant initiates an interaction with \newline
        a wild-type at location $i$. & \\
        \hline[dashed]
        A wild-type at location $i$ swaps with \newline
        a vacancy at location $i\pm 1$. & \SetCell[r=2]{m,7cm}\includegraphics[scale=0.25]{"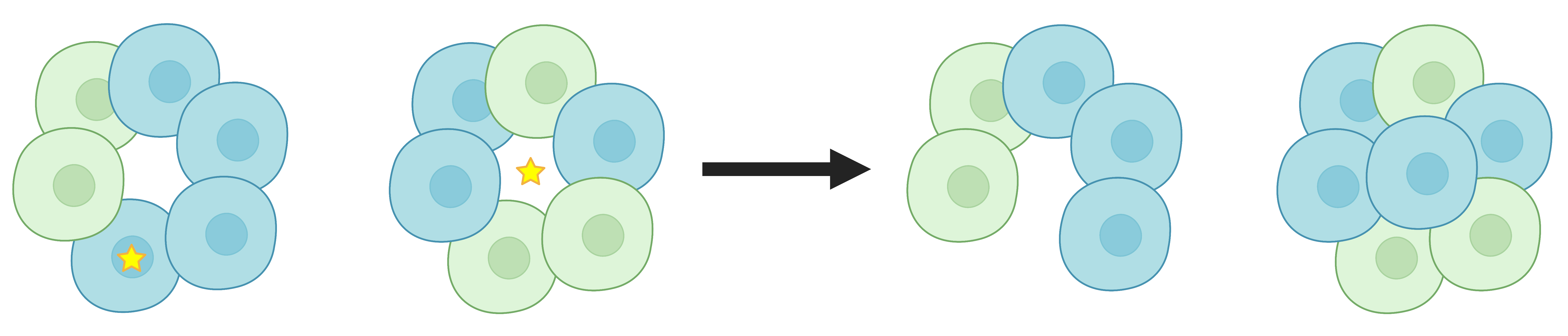"}\\
        \hline[dashed]
        A vacancy at location $i$ swaps with \newline
        a wild-type at location $i\pm 1$.&\\
        \hline[dashed]
        A mutant at location $i$ swaps with \newline
        a vacancy at location $i\pm 1$. & \SetCell[r=2]{m,7cm}\includegraphics[scale=0.25]{"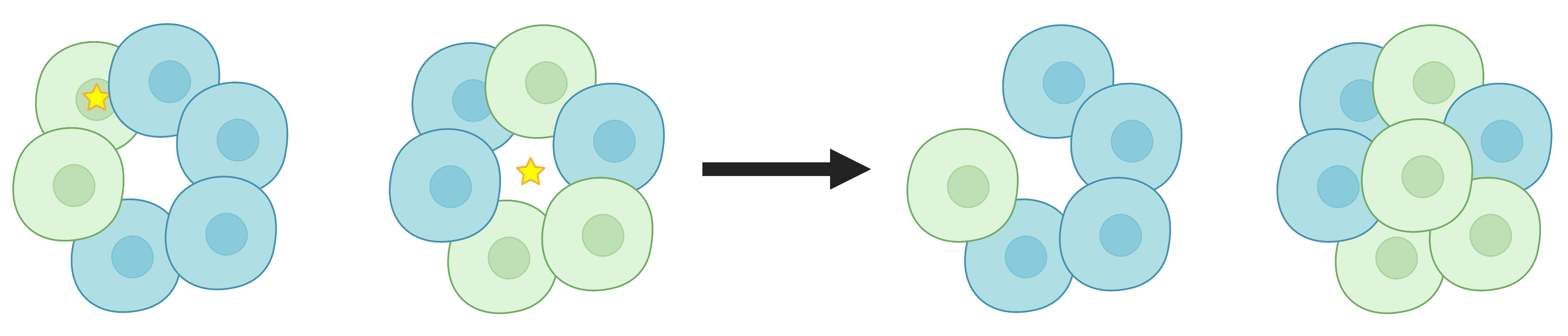"}\\
        \hline[dashed]
        A vacancy at location $i$ swaps with \newline
        a mutant at location $i\pm 1$. &\\
        \hline[dashed]
        A wild-type at location $i$ swaps with \newline
        a mutant at location $i\pm 1$. & \SetCell[r=2]{m,7cm}\includegraphics[scale=0.25]{"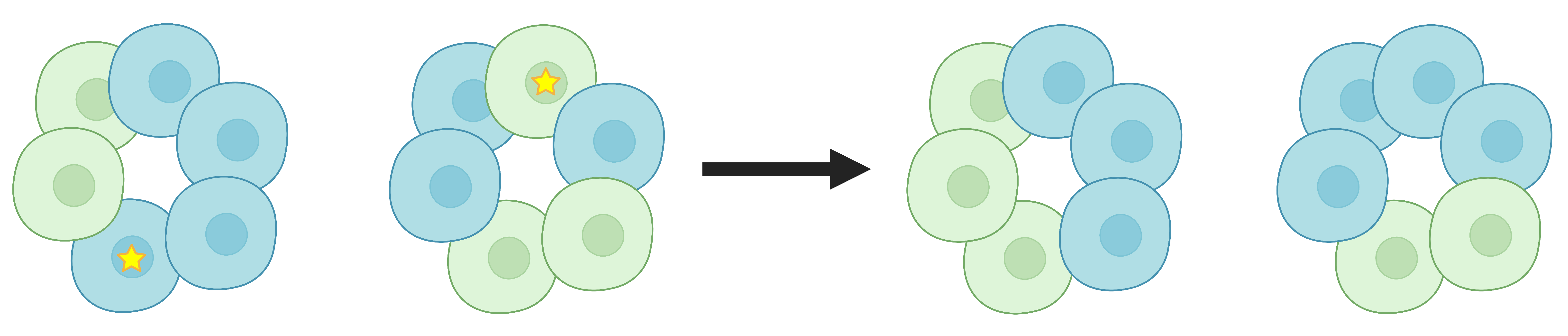"}\\
        \hline[dashed]
        A mutant at location $i$ swaps with \newline
        a wild-type at location $i\pm 1$. & \\
        \hline
    \end{tblr}
    
\end{table*}

\section{Stochastic Simulations}
We make use of the reaction framework described in the previous section, with an example given by Eq. (\ref{eq:reac1}). 
There, the magnitude of the payoff tensor elements determines the rates of each reaction, and the sign of the payoff tensor elements determine the stoichiometric components. 
In this way, higher-magnitude payoff tensor components lead to more frequent occurrences and a higher chance of a cell death or division depending on whether the payoff tensor component is positive or negative.
We maintain a finite carrying capacity ($C$) at each location by enforcing that if one cell has a positive payoff tensor component for a given interaction, the other cell has a negative payoff tensor component.
For example, if a wild-type interacts with a mutant and has a positive payoff tensor component for that interaction, then the mutant has a negative payoff tensor component for that interaction. 
In this instance the wild-type divides, the mutant dies, and the wild-type take over the spot of the mutant.
In our system, we have interactions between wild-types, mutants, and vacancies, which can all interact with each other in a local neighborhood called a deme.
Between demes, we allow for swapping events that incorporate the effects of diffusion.
All of these events are described in table (\ref{tab:events}).
In order to advance our simulations in time, we use the standard Gillespie algorithm \cite{gillespie1977exact}.

To initialize the system, we use an averaged wild-type wave profile $\langle w \rangle_{init}(x)$ from the results of $1000$ simulations of a system of only wild-types and vacancies. 
To generate this wave, we begin with a square wave of wild-types with a maximum at the carrying capacity.
Each wild-type wave develops into a sigmoidal shape, stochastically fluctuating around the average.
We align each stochastic wave at the half maximum value and take the average over all of the simulations.
To save computational memory, we set a barrier where upon being reached by a cell of the wild-type wave, a portion of the bulk of the wave is deleted and the entire wave is shifted back.
This mimics the notion of a comoving reference frame, and will become important for our discussion in the distinction between surfing and abiding as different types of survival.
We are interested in how likely it is for a mutant to survive if it arises at some position in this wild-type wave.
For a single simulation, we replace a wild-type at some position along the wave and allow the simulation to progress in time. 
We simulate $1000$ trajectories at each position and keep track of whether a mutant survived after $10$ million time steps, which was enough to see convergence of the survival probability.
Due to the nature of the comoving reference frame, it is possible that mutants that are left behind by the front will get deleted.
This effect misrepresents the true survival probability, only accounting for the contribution due to surfing. 
In order to account for this, we decouple the left and right portions of the simulation box once a cell has reached the barrier by deleting all wild-types, mutants, and vacancies between them.
We then have a comoving reference frame on the right portion of the box only.
This allows us to keep track of slow moving mutants in the bulk, while simultaneously keeping track of any that have managed to keep up with the front.

\begin{figure*}[ht]
  \centering
  \begin{minipage}[b]{0.49\textwidth}
    \includegraphics[width=\textwidth]{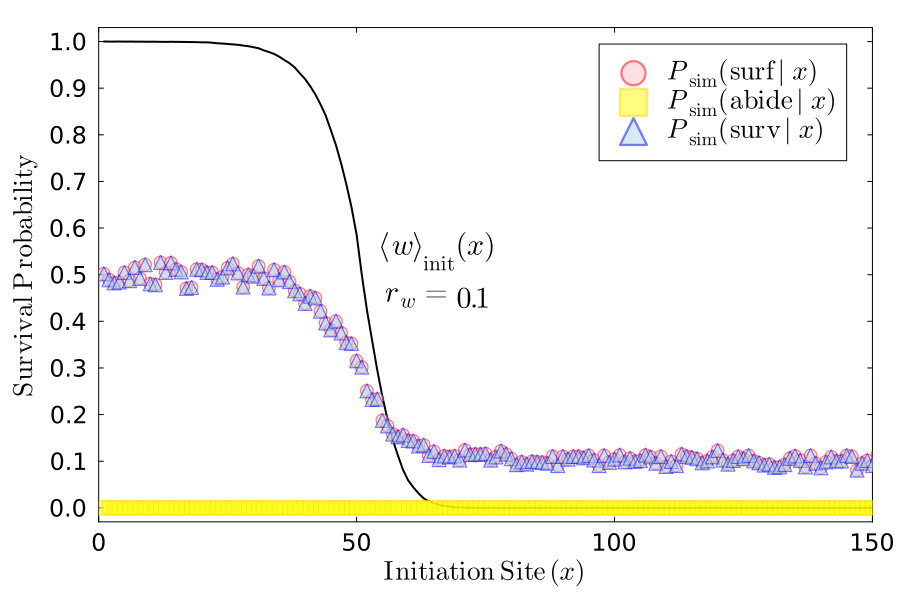}
  \end{minipage}
  \begin{minipage}[b]{0.49\textwidth}
    \includegraphics[width=\textwidth]{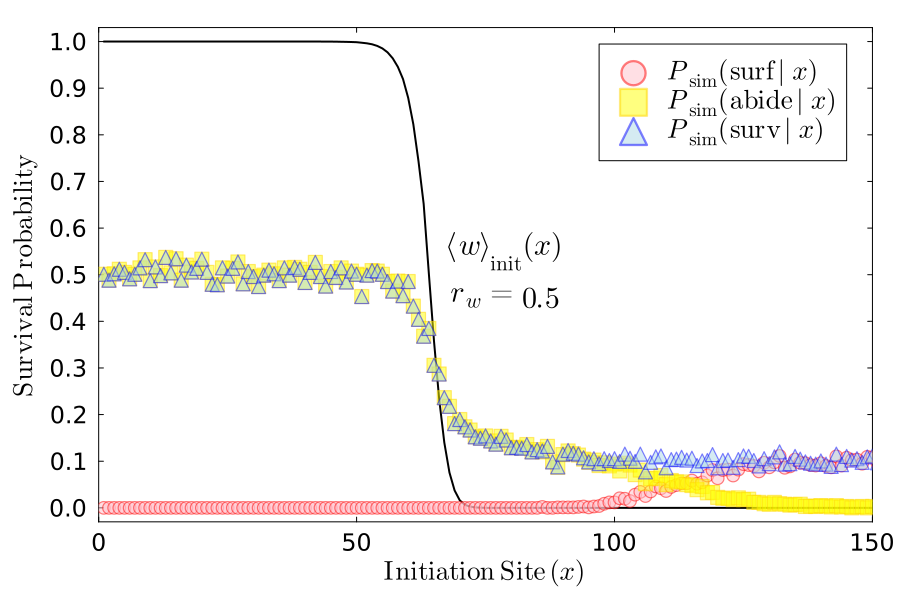}
  \end{minipage}
  \captionsetup{justification=raggedright}
  \caption{\textbf{Survival probability of a mutant can be split into surfing and abiding probabilities dependent on initiation position}. The surfing, abiding, and total survival probabilities are shown for a mutant initiated at a position $x$ where the asymmetric payoff tensor components are $P_{WM}^M=-0.25$ and $P_{MW}^M=0.5$ with a bulk mutant growth rate of $P_{WM}^M+P_{MW}^M=0.25$ and an intrinsic mutant growth rate of $r_m=0.1$. The black curve indicates the initial wild-type wave profile $\langle w \rangle _{init}(x)$ generated from $1000$ simulations of a system with only wild-types and vacancies. If the mutant is initiated in the bulk, it receives a boost in survival probability due to the net positive interaction with the wild-type which is greater than the mutant's intrinsic growth rate. (Left) The wild-type growth rate is given by $r_w=0.1$. In this case the mutant population is able to catch up to the front of the wave. In this instance, the total survival probability is comprised of pure surfing events. (Right) The wild-type growth rate is given by $r_w=0.5$. In this instance, many of the mutant populations are unable to keep up with the wild-type wave, and the total survival probability decomposes into a surfing portion and an abiding portion.}
  \label{fig:Surf_vs_Abide}
\end{figure*}

We can then account for both the surfing and abiding probabilities as they contribute to the total survival probability. 
We show an example of this decomposition in Fig. (\ref{fig:Surf_vs_Abide}).
For this example, we chose the asymmetric payoff tensor components to be $P_{WM}^M=-0.25$ and $P_{MW}^M=0.5$, corresponding to the wild-type and mutant interactions. 
We chose $r_m=0.1$, corresponding to the intrinsic growth rate of the mutant.
We vary the value of $r_w$, corresponding to the intrinsic growth rate of the wild-type to show the decomposition of the survival probability into a surfing portion and an abiding portion as $r_w$ increases.
An increase in the intrinsic wild-type growth rate has a corresponding increase in the wild-type wave speed, so we are effectively looking at the decomposition of the survival probability as the wild-type wave speed increases. 
For this choice of parameters, the mutant enjoys a benefit from being initiated in the bulk of the wave compared to being initiated at or ahead of the wave front. 
The net payoff is given by $P_{WM}^M+P_{MW}^M=0.25$ for a mutant interacting with a wild-type, corresponding to the growth rate of the mutant in the bulk. and is positively correlated with the wave speed of an established mutant population in the bulk.
We then see that for wild-type growth rates below $0.25$, the survival probability is determined by the surfing probability alone, as the mutant population is able to catch up to the wave front.
For wild-type growth rates above $0.25$, we see that the survival probability decomposes into surfing and abiding probabilities, as the mutant population may be left behind.
Similar phenomena have been observed in dimensions $2$, where homogeneous sectors of surfing cells or homogeneous bubbles of abiding cells form \cite{hallatschek2007genetic}. 
\section{Mathematical Model}
In order to develop a mathematical model for the survival probability of a new mutation in an expanding wild-type population, we must first make some assumptions. 
We begin with the probability, $p(x,t|\xi, \tau)$, that a mutation initiated at $\xi$ and time $\tau$, is later found at $x$ and time $t$, given the posterior probability that it survives.
We may then write the survival probability as the integral:
\begin{equation}
    u(\xi)=\int\limits_{-\infty}^\infty u(x)p(x, t|\xi,\tau)dx.
\end{equation}
We may then take the derivative of both sides with respect to $t$:
\begin{equation}
    0=\int\limits_{-\infty}^\infty u(x)\frac{dp(x, t|\xi,\tau)}{dt}dx.
\end{equation}
Our first approximation is that the derivative in the integral is given by Eq. (\ref{eq:det_ode}). Our second assumption is that the survival probability of a mutant is highly dependent on its ability to overcome the strong stochastic fluctuations early on.
This means that we may approximate the survival probability using the initial conditions of the system, those being the initial wild-type wave profile $\langle w \rangle_{init}(x)$. With these two assumptions, we may rewrite the derivative as follows:
\vspace{-1mm}
\begin{align}
    \frac{dp(x, t|\xi,\tau)}{dt}\approx r_mvm+(P_{WM}^M+P_{MW}^M)wm + \frac{d^2 m}{dx^2}\nonumber\\
    \approx r_m\langle v\rangle_{init}m+(P_{WM}^M+P_{MW}^M)\langle w\rangle_{init}m + \frac{d^2 m}{dx^2}\nonumber\\.
\end{align}
\vspace{-1mm}
This simply describes the effects of demography and dispersal when $t\approx\tau\ll1$. 
Plugging in the result we have:
\vspace{-1mm}
\begin{align}
    &\int\limits_{-\infty}^\infty u(x)\bigg(r_m\langle v\rangle_{init}m  \nonumber\\
    &+(P_{WM}^M+P_{MW}^M)\langle w\rangle_{init}m + \frac{d^2 m}{dx^2}\bigg)dx =0.
\end{align}
Using integration by parts, we may rewrite the integral as follows (see Eqs. B$5-$B$7$ in the appendix):
\begin{align}
    &\int\limits_{-\infty}^\infty m\bigg(r_m\langle v\rangle_{init}u\nonumber\\
    &+(P_{WM}^M+P_{MW}^M)\langle w\rangle_{init}u - \frac{d^2 u}{dx^2}\bigg)dx =0,
\end{align}
where we note that the derivatives of $u(x)$ should go to zero at the boundaries, and since $m(x,t)$ can be any arbitrary function, we have:
\begin{align}
        r_m\langle v\rangle_{init}u+(P_{WM}^M+P_{MW}^M)\langle w\rangle_{init}u - \frac{d^2 u}{dx^2}=0.
\end{align}
Further, we may add in the effect of a comoving reference frame with the wave velocity $s_{w,front}$:
\begin{align}
    &r_m\langle v\rangle_{init}u+(P_{WM}^M+P_{MW}^M)\langle w\rangle_{init}u \nonumber\\
    &- s_{w,front} \frac{du}{dx} - \frac{d^2 u}{dx^2}=0,
\end{align}
where we use the corresponding formula for the reference frame velocity:
\begin{equation}
    s_{w,front}= 2\sqrt{r_w}\Bigg(1-\frac{\pi^2}{2(\textrm{Ln}(C\sqrt{r_w})^2)}\Bigg),
\end{equation}
which is the leading-order expression for the wave speed of stochastic Fisher waves \cite{brunet1997shift}.
As in \cite{lehe2012rate}, we require a phenomenological term of order $u^2$ that accounts for the possibility of two mutants existing at short time scales, and additionally fixes the boundary conditions.
For our system, we have:
\begin{align}\label{eq:surv_ode}
         &r_m\langle v\rangle_{init}u+(P_{WM}^M+P_{MW}^M)\langle w\rangle_{init}u  \nonumber\\
        &- s_{w,front} \frac{du}{dx} - \frac{d^2 u}{dx^2} \nonumber\\ 
        &- [(\phi^+(P_{VM}^M)+\phi^+(P_{MV}^M))\langle v\rangle_{init}\nonumber\\
        &+(\phi^+(P_{WM}^M)+\phi^+(P_{MW}^M))\langle w\rangle_{init}]u^2=0.
\end{align}
\begin{figure*}[ht]
    \includegraphics[width=0.5\textwidth]{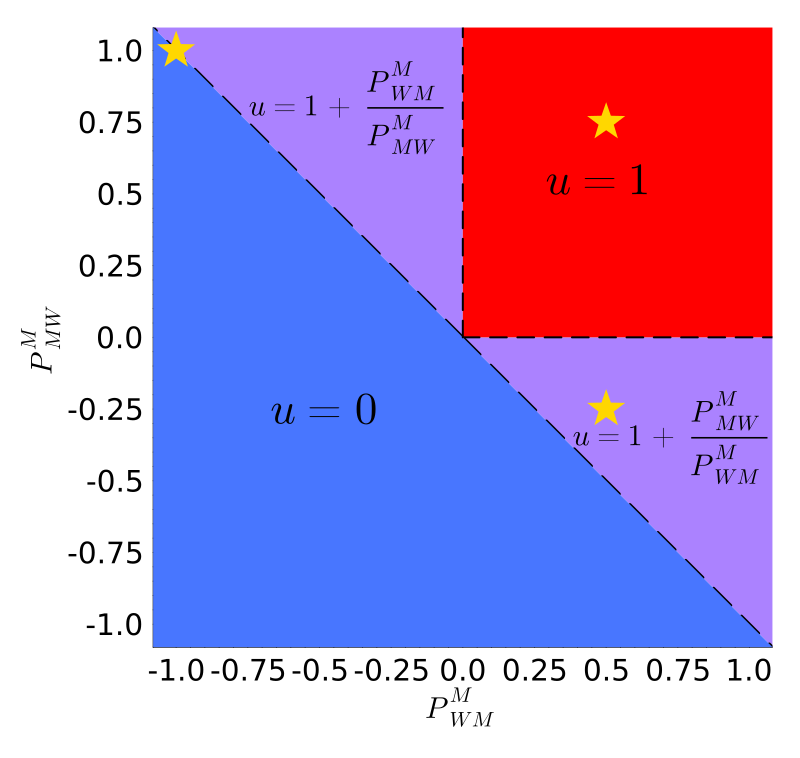}
  \captionsetup{justification=raggedright}
  \caption{\textbf{A phase diagram with analytical values of the survival probability of a mutant in the bulk of an expanding population }. The $x$ and $y$ axes are the two asymmetric payoff tensor components, which describe the payoff of a mutant interacting with a wild-type. In the region below $P_{MW}^M=-P_{WM}^M$, the survival probability is $u=0$, because the net payoff for a mutant interacting with a wild-type is negative. In the region between $P_{MW}^M=-P_{WM}^M$ and the positive unit square, the survival probability is between zero and unity while some of the interactions a mutant has with a wild-type are beneficial and some are detrimental, but the net payoff is positive. In the positive unit square, the survival probability of a mutant is equal to unity, because every interaction a mutant has with a wild-type is beneficial. Stars indicate the games, whose results are presented in this paper.}
  \label{fig:Clock_Plot}
\end{figure*}
One can find the full derivation in section B of the appendix. 
Here, $\phi^+(x)$ simply returns $x$ for $x \geq 0$.
Writing the equation in this way allows us to account for the full phase space of wild-type and mutant interactions.
For the boundary conditions we may use a Moran process as in \cite{nowak2006evolutionarych6} to find the survival probability of a single mutant.
We write the ratio of backwards to forwards transitions at $x\rightarrow -\infty$ and $x\rightarrow\infty$ as follows:
\begin{align}
    \gamma^{(-\infty)} &= \frac{\phi^-(P_{MW}^M)+ \phi^-(P_{WM}^M)}{\phi^+(P_{MW}^M)+ \phi^+(P_{WM}^M)}\nonumber\\
    \gamma^{(\infty)} &= \frac{\phi^-(P_{MV}^M)+ \phi^-(P_{VM}^M)}{\phi^+(P_{MV}^M)+ \phi^+(P_{VM}^M)},
\end{align}
where $\phi^-(x)$ simply returns $-x$ for $x \leq 0$. 
With these ratios we may write the boundary conditions as:
\begin{align}
    u(\pm\infty) = \lim\limits_{L\rightarrow\infty}\frac{1-\gamma^{(\pm\infty)}}{1-(\gamma^{(\pm\infty)})^{LC}}.
\end{align}
The phase diagram of the bulk boundary condition ($u(-\infty)$) is shown in Fig. (\ref{fig:Clock_Plot}).
For the example shown in Fig. (\ref{fig:Surf_vs_Abide}), We have that the mutant and wild-type interactions are given by $P_{WM}^M = -0.25$ and $P_{MW}^M = 0.5$, and the mutant and vacancy interactions are given by $P_{VM}^M=-(1-r_m)$ with $r_m=0.1$ and $P_{VM}^M=1$ .
We thus have that $\gamma^{(-\infty)}=0.5$ and $u(-\infty)=0.5$, and that $\gamma^{(\infty)}=0.9$ and $u(\infty)=0.1$ which recapitulates our results from the simulations.
\vspace{-4mm}
\section{Results}
\begin{figure*}
\centering
\subfloat[$P_{WM}^M=0.5,P_{MW}^M=0.75,r_w=0.1$]{
  \includegraphics[width=85mm]{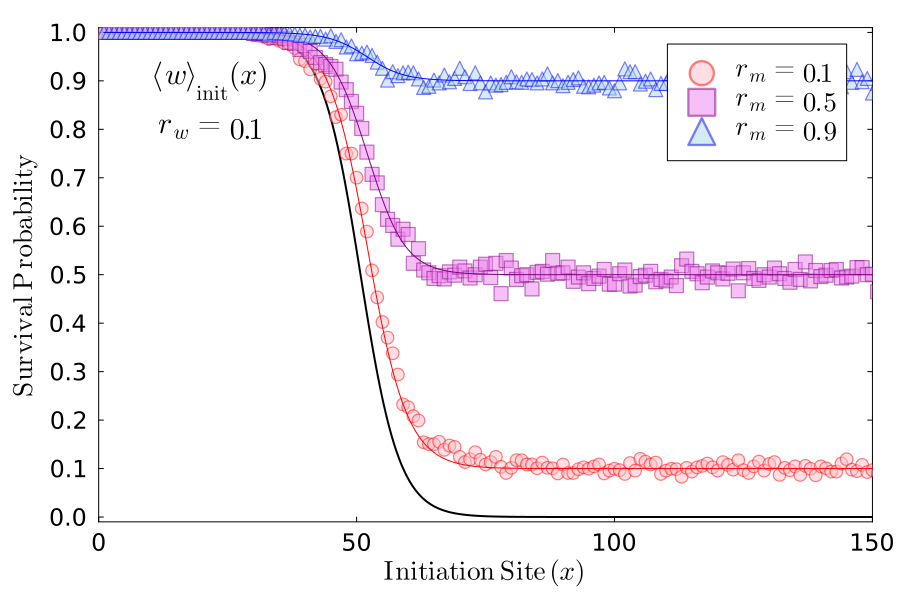}
}
\subfloat[$P_{WM}^M=0.5,P_{MW}^M=0.75,r_w=0.5$]{
  \includegraphics[width=85mm]{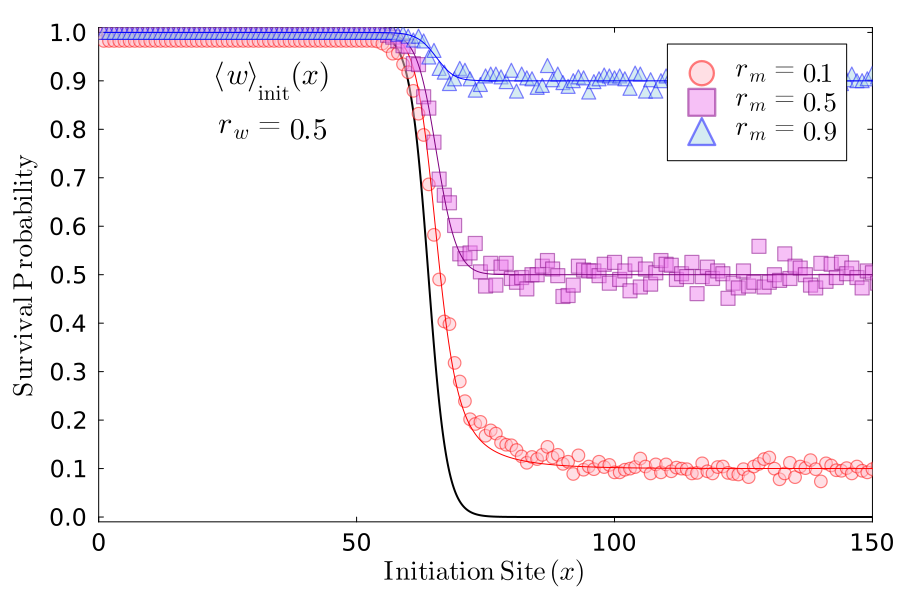}
}
\hspace{0mm}
\subfloat[$P_{WM}^M=0.5,P_{MW}^M=-0.25,r_w=0.1$]{
  \includegraphics[width=85mm]{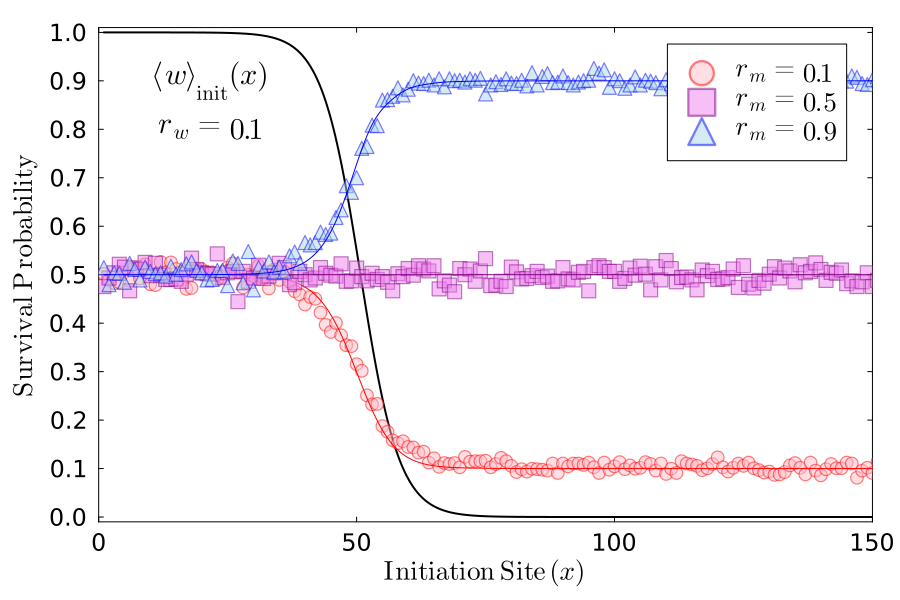}
}
\subfloat[$P_{WM}^M=0.5,P_{MW}^M=-0.25,r_w=0.5$]{
  \includegraphics[width=85mm]{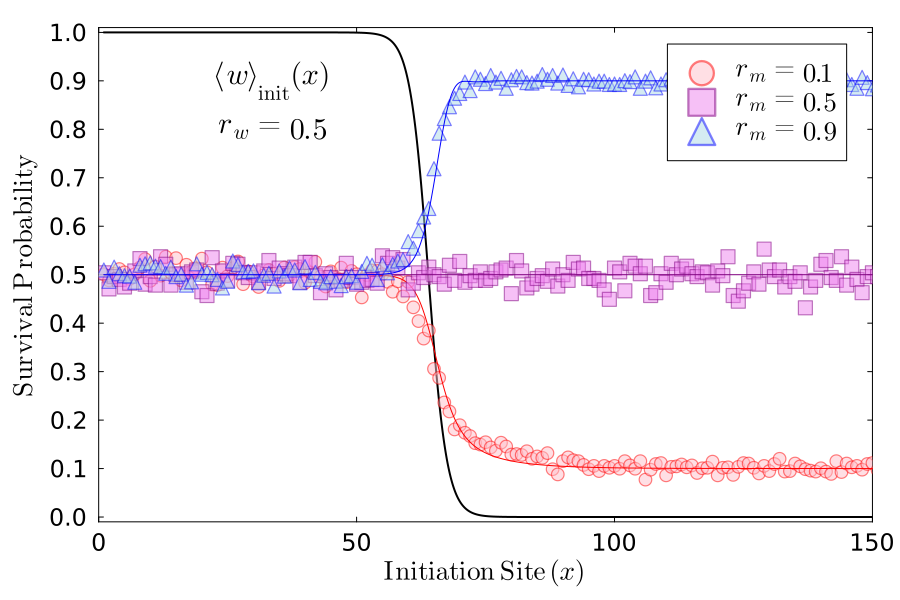}
}
\hspace{0mm}
\subfloat[$P_{WM}^M=-1.0,P_{MW}^M=1.0,r_w=0.1$]{ 
  \includegraphics[width=85mm]{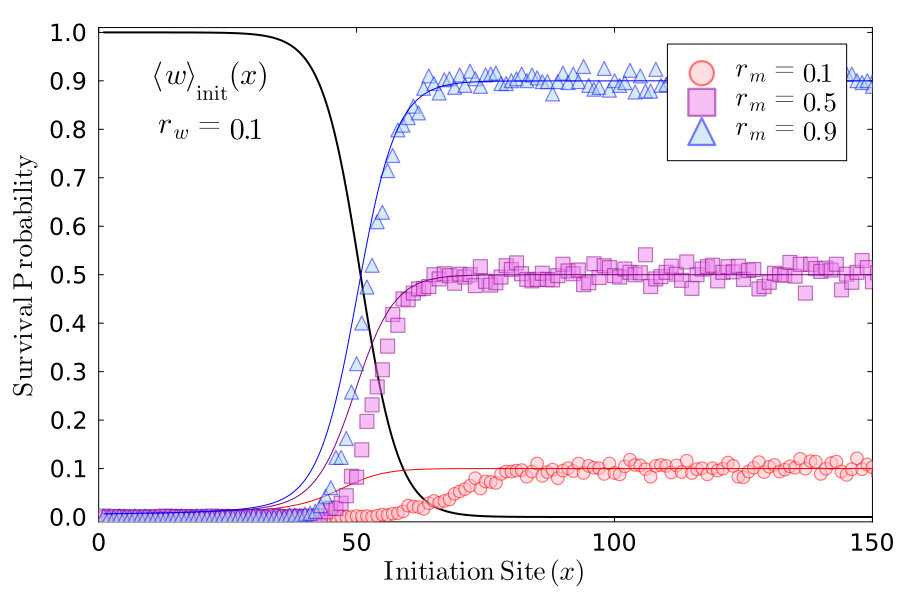}
}
\subfloat[$P_{WM}^M=-1.0,P_{MW}^M=1.0,r_w=0.5$]{
  \includegraphics[width=85mm]{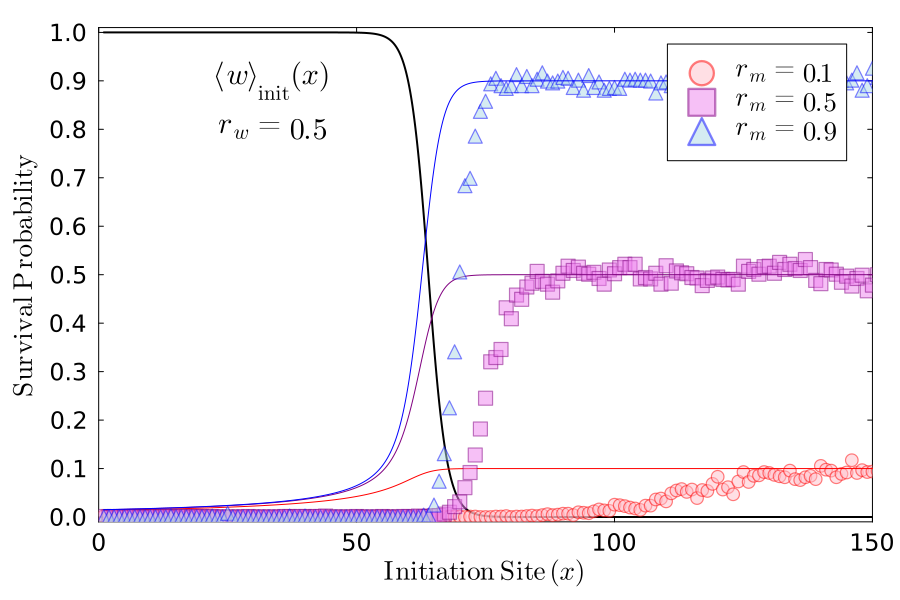}
}
\captionsetup{justification=raggedright}
\caption{\textbf{Comparison between simulations and our mathematical model of the survival probability across different asymmetric games and intrinsic growth rates}. Simulation results of the total survival probability given by scatter plot points, and the results of numerical integration of Eq. (\ref{eq:surv_ode}) for three sets of asymmetric game parameters with $r_w \in \{ 0.1,0.5\}$ and $r_m \in \{ 0.1,0.5,0.9\}$. The black curve indicates the initial wild-type wave profile $\langle w \rangle _{init}(x)$ generated from $1000$ simulations of a system with only wild-types and vacancies. (a) and (b):
For this game, the interaction between a wild-type and a mutant is always beneficial for the mutant, so the survival probability is equal to unity in the bulk.
 (c) and (d):
For this game, the interaction between a wild-type and a mutant is sometimes beneficial and sometimes detrimental for the mutant, so the survival probability is between zero and unity in the bulk.
 (e) and (f):
For this game, the interaction between a wild-type and a mutant is equally as beneficial as it is detrimental. Due to the large difference in initial population size between the mutant and the wild-type populations, the survival probability goes to zero in the bulk. 
Since the positive and negative contributions are equal in magnitude, the time for mutants to establish in the population is increased. 
When the bulk boundary condition is much smaller than the front boundary condition, an increase in establishment time leads to a discrepancy as our model uses the initial wild-type wave as input, thus over-predicting the number of vacancies a mutant will interact with. As the wild-type growth rate (wave velocity) increases, this discrepancy becomes larger.  
}
\label{fig:result_surv}
\end{figure*}

\vspace{-3mm}
We compare the results of simulations to the results of our mathematical model given by Eq. (\ref{eq:surv_ode}) in Fig. (\ref{fig:result_surv}). 
We choose three pairs of asymmetric game parameters, to display the regions of the phase space (stars in Fig. (\ref{fig:Clock_Plot})) in which there is agreement disagreement between the two.
We compare these games across a range of intrinsic growth rates of the wild-type ($r_w\in\{0.1,0.5\}$) and mutant ($r_m\in\{0.1,0.5,0.9\}$).
The first game (where $P_{WM}^M=0.5$ and $P_{MW}^M=0.75$) results in a positive payoff for mutants interacting with wild-types.
The result is a bulk survival probability of $u=1$, since there is no possibility for a mutant population to diminish in a wild-type environment.
The mutant is better off if it arises in the bulk for all of the values of the intrinsic growth rate we tested. 
In the absence of games, a mutant arising deep in the bulk
is very unlikely to establish a large population, but when supported by purely beneficial interactions, the mutant is guaranteed to survive in our model. 
Our mathematical model agrees very well with the results of our simulations in this region for a range of intrinsic growth rates of both the wild-type and the mutant.
The second game we compare is of a mixed nature, in that some of the interactions a mutant has with a wild-type are beneficial and some are detrimental.
The game parameters in this instance were $P_{WM}^M=0.5$ and $P_{MW}^M=-0.25$, resulting in a bulk survival probability of $u=0.5$.
In this instance, it may be either detrimental or beneficial for a mutant to arise in the bulk, depending on its intrinsic growth rate. 
Again, our model agrees quite well with the results of the simulations for the parameter ranges we tested.
The third game we compare is a game where the payoff tensor components are equal in magnitude but opposite in sign, thus having equally beneficial and detrimental interactions for the mutant. 
For this game, we chose  $P_{WM}^M=-1.0$ and $P_{MW}^M=1.0$.
When the detrimental interactions are greater or equal in magnitude to the beneficial interactions, the time for a mutant to establish a colony becomes large.
The larger the ratio of the bulk boundary condition to the front boundary condition ($u_{-\infty}/u_{\infty}$), the more of a discrepancy we will observe between our mathematical model and simulation results. 
This is attributed to the assumption of the short time limit in our mathematical model, which over-predicts the number of vacancies a mutant will interact with.
This difference is further enhanced by larger wild-type growth rates (wave speeds).
Our model is valid when the ratio ($u_{-\infty}/u_{\infty} \ge 1$) for all wild-type and mutant growth rates we tested. For ($u_{-\infty}/u_{\infty} < 1$), the model becomes more accurate at lower wild-type growth rates and at higher ratios. 
Our model is valid in the interesting regimes where the mutant is less fit on its own, but receives a relative increase in fitness from interacting with wild-types.
These interactions are especially important for the case when the mutant fitness decrease is due to some cost to developing drug resistance.
In the absence of treatment, it is possible that extant resistance mutants may survive by either abiding or surfing.
Once treatment resumes, an advantage of a higher starting frequency (in addition to its higher fitness compared to the wild-type in treatment) may lead to both an increase in fixation probability and a decrease in the fixation time.
Both of these modifications are detrimental in cancer and bacterial infection progression.
Our model is approximately valid for the case where a mutant is more fit on its own, and the wild-type wave is moving slowly.
As the intrinsic growth rate of the mutant increases, this approximation becomes more accurate.
In contrast to the other case, we may consider a situation where the initial wild-type is resistant to a dr ug, and the mutant is sensitive to a drug, and depending on the fitness decrease experienced upon receiving treatment, it may still be possible that a mutant has a positive survival probability further towards the front of a wild-type wave.
\section{Discussion}
In this paper, we demonstrate that it is possible to predict the probability of survival of a mutant in an expanding population, depending on ($1$) the location where the mutant first appeared in the population and ($2$) the types of interactions the mutant experiences with other cells in its environment. 
This extends the existing literature on range expansion that has previously focused on neutral mutations or specific types of interactions to a more general framework.
In many models that include ecological interactions, spatial structure is ignored, and in many spatial models, ecological interactions are ignored. 
Our model describes the full three-tiered framework where spatial structure modifies the effect of stochasticity on ecological interactions, and ecological interactions modify intrinsic fitness.
We show how these effects combine to influence the survival probability of new mutations.
By combining spatial density and frequency dependence into our models one may better predict ($1$) where a mutant is likely to establish, ($2$) if the interactions it experiences with other cells support or hamper its growth, and ($3$) if the interactions support the development of heterogeneity within the population.
Genetic heterogeneity is one of the main determiners of the development of drug resistance in cancer, bacterial infections, and viral infections \cite{nikaido2009multidrug,poole2001multidrug,persidis1999cancer,gillet2010mechanisms,szakacs2006targeting,rojas2022genomic}.
In order to inform better treatment strategies, better models of the spatial distribution of mutant survival and fitness need to be developed.
Given the growing experimental evidence for significant ecological interactions within pathologically expanding populations (such as cancer \cite{farrokhian2022measuring,maltas2024frequency,
nichol2015steering,weaver2024reinforcement,lin2023evolution,iram2021controlling,lassig2023steering,west2023survey} and bacterial systems \cite{hibbing2010bacterial,burmolle2014interactions,stein2013ecological}), this theoretical framework is required to model such effects. 
The limitations of our modeling framework include the restriction to a one-dimensional system.
While the mathematical model may be extended in a straightforward way to higher dimensions, computational simulations in higher dimensions are a limiting factor. 
We restrict our analysis to the off-diagonal elements of the payoff tensor, but in principal the parameter ranges we explore could be extended as well. 
This model could be extended by allowing for interactions between neighboring locations, or may be extended to even larger interaction distances.
The number of mutations could be extended as well, but many of the assumptions we make in our model no longer apply, thus making calculations significantly more cumbersome.
One of the more straightforward ways to extend our model is by allowing a concentration gradient of a drug, such as an antibiotic or chemotherapy.
With this addition, one could observe how treatment may influence the development of resistance mutations.
In the future, more detailed analysis of the fitness profile as it changes over time would be a useful addition to the study of pushed and pulled waves.
Pushed/pulled waves exhibit higher/lower fitness in the bulk than at the front and often support higher/lower heterogeneity \cite{hallatschek2008gene,gandhi2016range, peischl2015expansion, legault2020interspecific, hallatschek2008gene,roques2012allee}.
Finally, our model can be parameterized using experimental and in-vivo data to infer the likelihood of survival depending on where a mutation arises.
In combination with an extended model which includes how drugs affect ecological interactions, our model could infer treatment strategies which mitigate the risk of nurturing drug resistant mutations.

\section{Author Contributions}

\textbf{Jason M. Gray} Conceptualization, Methodology, Formal Analysis, Software, Writing - Original Draft, Writing - Review \& Editing, Visualization, Investigation, Validation.

\textbf{Rowan Barker-Clarke} Conceptualization, Writing - Review \& Editing.

\textbf{Jacob G. Scott} Conceptualization, Methodology, Formal Analysis, Writing - Review \& Editing, Project administration, Funding acquisition,Supervision.

\textbf{Michael Hinczewski} Conceptualization, Methodology, Formal Analysis, Writing - Review \& Editing, Project administration.

Corresponding Author (jmg367@case.edu, grayjason13@gmail.com).

\section{Acknowledgements}

The authors are grateful for insightful discussions with Dr. Oskar Hallatschek. The authors also thank Case Western Reserve University for providing High Performance Computing Cluster (HPCC) resources. 
Overall support for this project was provided by (NIH R37CA244613).

\clearpage
\newpage

\pagebreak
\nocite{*}
\bibliographystyle{unsrt}
\bibliography{apssamp}% Produces the bibliography via BibTeX.

\pagebreak
\section*{Appendices}
\appendix

\section{Reaction-Diffusion Equations with Asymmetric Games}
In the system we consider, there are three types which interact: wild-types, mutants, and vacancies. To make the distinction between numerical values and indices we define an index set $\mathcal{I}=\{W_1,M_1,V_1,\dots,W_L,M_L,V_L\}$ which has a length of $3L$, where $L$ is the number of adjascent locations. At each location (denoted by $i$), the number of each type is given by $N_{W_i}$, $N_{M_i}$, and $N_{V_i}=C-N_{W_i}-N_{M_i}$ respectively. Here, $C$ is the carrying capacity of a single location. The frequencies of each type is given by $w_i=N_{W_i}/C$, $m_i=N_{M_i}/C$, and $v_i=N_{V_i}/C=1-w_i-m_i$ respectively. The state vector ($\vec{x}$) of the system has a length of $3L$, and can be written as:

\begin{equation}
    \vec{x}=(w_1,m_1,v_1,\dots,w_L,m_L,v_L)
\end{equation}

Interactions between different types within a given location can be summarized with an asymmetric payoff tensor ($P$) which is indexed by the index set $\mathcal{I}$. We only consider interactions within locations, not between them. We thus have a block diagonal asymmetric payoff tensor. We emphasize a block corresponding to a single deme is as follows:

\begin{widetext}
\begin{equation*}
    P=\begin{pmatrix}
        (P_{W_1W_1}^{W_1},P_{W_1W_1}^{M_1},P_{W_1W_1}^{V_1},\dots,0)&(P_{W_1M_1}^{W_1},P_{W_1M_1}^{M_1},P_{W_1M_1}^{V_1},\dots,0)&(P_{W_1V_1}^{W_1},P_{W_1V_1}^{M_1},P_{W_1V_1}^{V_1},\dots,0) & & &\\
        (P_{M_1W_1}^{W_1},P_{M_1W_1}^{M_1},P_{M_1W_1}^{V_1},\dots,0)&(P_{M_1M_1}^{W_1},P_{M_1M_1}^{M_1},P_{M_1M_1}^{V_1},\dots,0)&(P_{M_1V_1}^{W_1},P_{M_1V_1}^{M_1},P_{M_1V_1}^{V_1},\dots,0) & & \dots & \mathbf{0}\\
        (P_{V_1W_1}^{W_1},P_{V_1W_1}^{M_1},P_{V_1W_1}^{V_1},\dots,0)&(P_{V_1M_1}^{W_1},P_{V_1M_1}^{M_1},P_{V_1M_1}^{V_1},\dots,0)&(P_{V_1V_1}^{W_1},P_{V_1V_1}^{M_1},P_{V_1V_1}^{V_1},\dots,0) & & &\\
        & \vdots &  & \ddots\\
        & \mathbf{0} & & 
    \end{pmatrix}.
\end{equation*}
\end{widetext}

Here, $P_{AB}^Z$ corresponds to the expected payoff for an individual of type $Z$ when an individual of type $A$ initiates an interaction and an individual of type $B$ accepts the interaction. In other words, $A$ is player $1$ in the game and $B$ is player $2$. ``$\mathbf{0}$'' is a $(3,3,3L)$-tensor with zeros for every element. In our system, we do not consider ``self-interactions'', so we require that $(\forall A)[P_{AA}^Z=0]$. We also, do not consider payoffs for individuals which are not initiating or accepting a given interaction, so we require that $(\forall A,\forall B)[P_{AB}^Z=0|Z \ne A \land Z \ne B]$. We thus have a payoff tensor of the form:

\begin{widetext}
\begin{equation}
    P=\begin{pmatrix}
       \vec{0}&(P_{W_1M_1}^{W_1},P_{W_1M_1}^{M_1},0,\dots,0)&(P_{W_1V_1}^{W_1},0,P_{W_1V_1}^{V_1},\dots,0) & & &\\
        (P_{M_1W_1}^{W_1},P_{M_1W_1}^{M_1},0,\dots,0)&\vec{0}&(0,P_{M_1V_1}^{M_1},P_{M_1V_1}^{V_1},\dots,0) & & \dots & \mathbf{0}\\
        (P_{V_1W_1}^{W_1},0,P_{V_1W_1}^{V_1},\dots,0)&(0,P_{V_1M_1}^{M_1},P_{V_1M_1}^{V_1},\dots,0)&\vec{0} & & &\\
        & \vdots &  & \ddots\\
        & \mathbf{0} & & 
    \end{pmatrix}.
\end{equation}
\end{widetext}

We assume uniformity across space, meaning that $(\forall A,\forall B,\forall Z,\forall i,\forall j)[P_{A_iB_i}^Z=P_{A_jB_j}^Z]$. From now on, we will write $P_{A_iB_i}^{Z_i}=P_{AB}^Z$ for brevity. To provide an example of how one would use this payoff tensor, we consider the case when a wild-type initiates and interaction with a mutant in the same deme ($i$), and we would like to know how this interaction affects the number of wild-types for the chosen deme. To define this mathematically, we make use of the propensity vector ($\vec{\rho}_{AB}(P)$). The propensity vector for this reaction may be written in the following way:

\begin{equation}
    \vec{\rho}_{W_iM_i}(P)=\begin{pmatrix}
        \vdots\\
        |P_{WM}^{W}| w_im_i\\
        |P_{WM}^{M}| w_im_i\\
        0\\
        \vdots
    \end{pmatrix}.
\end{equation}

The notation $|\cdot|$ is the absolute value. The propensity function must be positive by definition as it is essentially an unnormalized probability of a given reaction. The remaining terms $w_i$ and $m_i$ are the frequencies, which can be interpreted as the probability to select a wild-type in deme $i$ and a mutant in deme $i$. 

Each reaction has a corresponding ``state change vector'' ($\vec{\nu}_{AB}(P)$) which gives us the amount each species changes by in a given reaction.  For the example above, we have:

\begin{equation}
\begin{aligned}
     \vec{\nu}_{W_iM_i}(P) = \begin{pmatrix}
         \vdots\\
         \textrm{sgn}(P_{WM}^{W})\\
         \textrm{sgn}(P_{WM}^{M})\\
         0\\
         \vdots
     \end{pmatrix},
\end{aligned}
\end{equation}

where $\textrm{sgn}(x)$ is the sign of $x$.

In addition to interactions between individuals, we also allow for motility (\emph{i.e.} swapping) between neighboring demes. Motility can be viewed as a $4$-body interaction. For example, if a wild-type in deme $i$ were to swap places with a vacancy in deme $i+1$, then the number of wild-types in deme $i$ would go down by one ($W_i \rightarrow W_i-1$), the number of wild-types in deme $i+1$ would go up by one ($W_{i+1} \rightarrow W_{i+1}+1$), 
the number of vacancies in deme $i$ would go up by one ($V_{i} \rightarrow V_{i}+1$), 
and the number of vacancies in deme $i+1$ would go down by one ($V_{i+1} \rightarrow V_{i+1}-1$).

We may condense the swapping reactions into a tensor ($S$), which is similar to the expected payoff tensor in form. In this case, we do allow for ``payoffs'' where an individual is not involved in an initiator-acceptor pair. Additionally, we assume a finite space and a spatially uniform swapping tensor except at the boundaries. We assume symmetric swapping rates between left and right so that the swapping tensor components have the following form away from the boundaries:

\noindent\begin{minipage}{.5\linewidth}
\begin{equation*}
  \begin{aligned}
       S_{A_iB_{i\pm1}}^{A_i}&=-\frac{D}{2(\Delta x)^2}\\
    S_{A_iB_{i\pm1}}^{B_{i\pm1}}&=-\frac{D}{2(\Delta x)^2}\\
    S_{A_iB_{i\pm1}}^{A_{i\pm1}}&=\frac{D}{2(\Delta x)^2}\\
    S_{A_iB_{i\pm1}}^{B_i}&=\frac{D}{2(\Delta x)^2}\\
    \vdots
\end{aligned}
\end{equation*}
\end{minipage}%
\begin{minipage}{.5\linewidth}
\begin{equation}
  \begin{aligned}
       S_{B_{i\pm1}A_i}^{A_i}&=-\frac{D}{2(\Delta x)^2}\\
    S_{B_{i\pm1}A_i}^{B_{i\pm1}}&=-\frac{D}{2(\Delta x)^2}\\
    S_{B_{i\pm1}A_i}^{A_{i\pm1}}&=\frac{D}{2(\Delta x)^2}\\
    S_{B_{i\pm1}A_i}^{B_i}&=\frac{D}{2(\Delta x)^2}\\
    \vdots
\end{aligned}
\end{equation}
\end{minipage}

where $\Delta x$ represents the distance between neighboring locations, and $D$ represents the diffusion constant with units of squared distance over time $[d]^2[t]^{-1}$.
There are eight additional components, with the initiator-acceptor pairs being comprised of $A_{i\pm1}$ and $B_i$. 
We assume that $A$ and $B$ are not the same type since swapping of equal types maintains the state vector, effectively lengthening the time scale. 
Since we are only interested in survival probabilities, and not in fixation (surfing) times, this is a valid assumption. 
All components of the swapping tensor which do not follow this relationship between the indices are assumed to be zero. We may define a propensity for swapping reactions as well. 
As an example, we consider a wild-type in deme $i$ swapping with a vacancy in deme $i+1$. 
The propensity vector for this reaction would be:

\begin{equation}
\begin{aligned}
      \vec{\rho}_{W_iV_{i+1}}(S)&=
    \begin{pmatrix}
        \vdots\\
        |S_{W_iV_{i+1}}^{W_i}| w_iv_{i+1}\\
        0\\
        |S_{W_iV_{i+1}}^{V_i}| w_iv_{i+1}\\
        |S_{W_iV_{i+1}}^{W_{i+1}}| w_iv_{i+1}\\
        0\\
        |S_{W_iV_{i+1}}^{V_{i+1}}| w_iv_{i+1}\\
        \vdots
    \end{pmatrix}  \\
    &=
    \begin{pmatrix}
        \vdots\\
        \frac{D}{2(\Delta x)^2} w_iv_{i+1}\\
        0\\ 
        \frac{D}{2(\Delta x)^2} w_iv_{i+1}\\
        \frac{D}{2(\Delta x)^2} w_iv_{i+1}\\
        0\\ 
        \frac{D}{2(\Delta x)^2} w_iv_{i+1}\\
        \vdots
    \end{pmatrix}.
\end{aligned}
\end{equation}

We may define a corresponding state change vector as we had for the payoff tensor interactions. For the example above, we have:

\begin{equation}
\begin{aligned}
     \vec{\nu}_{W_iV_{i+1}}(S) = \begin{pmatrix}
         \vdots\\
         \textrm{sgn}(S_{W_iV_{i+1}}^{W_i})\\
         0\\
         \textrm{sgn}(S_{W_iV_{i+1}}^{V_i})\\
         \textrm{sgn}(S_{W_iV_{i+1}}^{W_{i+1}})\\
         0\\
         \textrm{sgn}(S_{W_iV_{i+1}}^{V_{i+1}})\\
         \vdots
     \end{pmatrix}.
\end{aligned}
\end{equation}

We may combine our payoff tensor and swapping tensor into a ``reaction-diffusion'' tensor, defined simply as $R=P+S$. The propensity function evaluated with the reaction-diffusion tensor is as follows:

\begin{equation}\label{eq:full-prop}
\begin{aligned}
      \vec{\rho}_{\mathcal{I}_i\mathcal{I}_j}(R)&=
    \begin{pmatrix}
        |R_{\mathcal{I}_i\mathcal{I}_j}^{W_1}|x_ix_j\\
        |R_{\mathcal{I}_i\mathcal{I}_j}^{M_1}|x_ix_j\\
        |R_{\mathcal{I}_i\mathcal{I}_j}^{V_1}|x_ix_j\\
        \vdots\\
        |R_{\mathcal{I}_i\mathcal{I}_j}^{W_L}|x_ix_j\\
        |R_{\mathcal{I}_i\mathcal{I}_j}^{M_L}|x_ix_j\\
        |R_{\mathcal{I}_i\mathcal{I}_j}^{V_L}|x_ix_j\\
    \end{pmatrix}.  
\end{aligned}
\end{equation}

Likewise, our state-change vector is given by:

\begin{equation}\label{eq:state-change-vecs}
\begin{aligned}
      \vec{\nu}_{\mathcal{I}_i\mathcal{I}_j}(R)&=
    \begin{pmatrix}
        \textrm{sgn}(R_{\mathcal{I}_i\mathcal{I}_j}^{W_1})\\
        \textrm{sgn}(R_{\mathcal{I}_i\mathcal{I}_j}^{M_1})\\
        \textrm{sgn}(R_{\mathcal{I}_i\mathcal{I}_j}^{V_1})\\
        \vdots\\
        \textrm{sgn}(R_{\mathcal{I}_i\mathcal{I}_j}^{W_L})\\
        \textrm{sgn}(R_{\mathcal{I}_i\mathcal{I}_j}^{M_L})\\
        \textrm{sgn}(R_{\mathcal{I}_i\mathcal{I}_j}^{V_L})\\
    \end{pmatrix}.
\end{aligned}
\end{equation}

Our deterministic dynamics are then given by:

\begin{equation}
    \begin{aligned}
        \frac{d\vec{x}}{d t} = \mathlarger{\sum}\limits_{A\in \mathcal{I}}\mathlarger{\sum}\limits_{B\in \mathcal{I}}\vec{\nu}_{AB}(R)\odot\vec{\rho}_{AB}(R),
    \end{aligned}
\end{equation}

where $\odot$ is the ``element-wise'' or ``Hadamard'' product. Due to the assumptions of only considering in-deme interactions, nearest-neighbor swapping, and no self-interactions, our dynamical equations simplify in the following way (for a single element):

\begin{equation}\label{eq:app_det_ode_mut}
    \begin{aligned}
        \frac{d x_i}{d t} = \mathlarger{\sum}\limits_{j=i-1}^{i+1}\mathlarger{\sum}\limits_{k=i-1}^{i+1}\mathop{\mathlarger{\sum}\limits_{A\in \{W_j,M_j,V_j\}}}_{B\in \{W_k,M_k,V_k\}}\nu_{AB}^{\mathcal{I}_i}(R)\rho_{AB}^{\mathcal{I}_i}(R).
    \end{aligned}
\end{equation}

As an example, we will calculate the dynamics for a mutant in deme $i$ which is away from the boundaries of the system. The only nonzero payoffs of the form $R_{AB}^{M_i}$ are in the submatrix corresponding to the $W_{i-1}$ to $V_{i+1}$ rows and columns. This is because we only consider up to nearest neighbor interactions. The sub matrix is as follows:
\begin{widetext}
\begin{equation*}\label{eq:Block-m-matrix}
    R_{sub}^{M_i}=
    \begin{pmatrix}
        0 &            0 &           0 &            0 &           -\frac{D}{2(\Delta x)^2} & 0 &           0 &            0 &           0\\
        0 &            0 &           0 &            \frac{D}{2(\Delta x)^2} & 0 &            \frac{D}{2(\Delta x)^2} & 0 &            0 &           0\\
        0 &            0 &           0 &            0 &           -\frac{D}{2(\Delta x)^2} & 0 &           0 &            0 &           0\\
        0 &            \frac{D}{2(\Delta x)^2} & 0 &            0 &           P_{WM}^M &     0 &           0 &            \frac{D}{2(\Delta x)^2} & 0\\
        -\frac{D}{2(\Delta x)^2} & 0 &           -\frac{D}{2(\Delta x)^2} & P_{MW}^M &    0 &            P_{MV}^M &    -\frac{D}{2(\Delta x)^2} & 0 &           -\frac{D}{2(\Delta x)^2}\\
        0 &            \frac{D}{2(\Delta x)^2} & 0 &            0 &           P_{VM}^M &     0 &           0 &            \frac{D}{2(\Delta x)^2} & 0\\
        0 &            0 &           0 &            0 &           -\frac{D}{2(\Delta x)^2} & 0 &           0 &            0 &           0\\
        0 &            0 &           0 &            \frac{D}{2(\Delta x)^2} & 0 &            \frac{D}{2(\Delta x)^2} & 0 &            0 &           0\\
        0 &            0 &           0 &            0 &           -\frac{D}{2(\Delta x)^2} & 0 &           0 &            0 &           0
    \end{pmatrix}.
\end{equation*}
\end{widetext}
% \clearpage
% \newpage
% \mbox{~}
% \clearpage
% \newpage
With this, we have twenty propensity vector components ($\rho_{AB}^{M_i}$) and twenty state-change vector components ($\nu_{AB}^{M_i}$):

\begingroup
\renewcommand{\arraystretch}{2} % Default value: 1
    \begin{table}[!ht]
        \centering
        \begin{tabular}{ |l|l| }
               \hline
               \textit{Propensity} & \textit{State-change Vector}\\
               \hline
               $\rho_{M_iV_i}^{M_i}=|P_{MV}^M|m_iv_i $ & $\nu_{M_iV_i}^{M_i}=\textrm{sgn}(P_{MV}^M)$\\
               $\rho_{M_iW_i}^{M_i}=|P_{MW}^M|m_iw_i $ & $\nu_{M_iW_i}^{M_i}=\textrm{sgn}(P_{MW}^M)$\\
               $\rho_{V_iM_i}^{M_i}=|P_{VM}^M|m_iv_i $ & $\nu_{V_iM_i}^{M_i}=\textrm{sgn}(P_{VM}^M)$\\
               $\rho_{W_iM_i}^{M_i}=|P_{WM}^M|m_iw_i $ & $\nu_{W_iM_i}^{M_i}=\textrm{sgn}(P_{WM}^M)$\\
               $\rho_{M_iV_{i\pm1}}^{M_i}=(\frac{D}{2(\Delta x)^2})m_iv_{i\pm1} $ & $\nu_{M_iV_{i\pm1}}^{M_i}=-1$\\
               $\rho_{M_iW_{i\pm1}}^{M_i}=(\frac{D}{2(\Delta x)^2})m_iw_{i\pm1} $ & $\nu_{M_iW_{i\pm1}}^{M_i}=-1$\\
               $\rho_{V_{i\pm1}M_i}^{M_i}=(\frac{D}{2(\Delta x)^2})m_iv_{i\pm1} $ & $\nu_{V_{i\pm1}M_i}^{M_i}=-1$\\
               $\rho_{W_{i\pm1}M_i}^{M_i}=(\frac{D}{2(\Delta x)^2})m_iw_{i\pm1} $ & $\nu_{W_{i\pm1}M_i}^{M_i}=-1$\\
               $\rho_{M_{i\pm1}V_i}^{M_i}=(\frac{D}{2(\Delta x)^2})m_{i\pm1}v_i $ & $\nu_{M_{i\pm1}V_i}^{M_i}=1$\\
               $\rho_{M_{i\pm1}W_i}^{M_i}=(\frac{D}{2(\Delta x)^2})m_{i\pm1}w_i $ & $\nu_{M_{i\pm1}W_i}^{M_i}=1$\\
               $\rho_{V_iM_{i\pm1}}^{M_i}=(\frac{D}{2(\Delta x)^2})m_{i\pm1}v_i $ & $\nu_{V_iM_{i\pm1}}^{M_i}=1$\\
               $\rho_{W_iM_{i\pm1}}^{M_i}=(\frac{D}{2(\Delta x)^2})m_{i\pm1}w_i $ & $\nu_{W_iM_{i\pm1}}^{M_i}=1$\\
               \hline
        \end{tabular}
        \captionsetup{justification=raggedright}
        \caption{Propensities and their corresponding stoichiometric values for events that effect the mutant at location $i$ ($M_i$).}
        \label{tab:propensities_and_changevec}
    \end{table}
\endgroup
According to Eq. (\ref{eq:app_det_ode_mut}), noting that $\textrm{sgn}(x)|x|=x$,we have:

\begin{align}
    \frac{d m_i}{d t} &= 
    P_{MV}^Mm_iv_i+P_{VM}^Mv_im_i+ P_{MW}^Mm_iw_i+P_{WM}^Mw_im_i\nonumber\\
 &- \frac{D}{2(\Delta x)^2}\Bigg[m_iv_{i+1} + m_iv_{i-1}+ m_iw_{i+1}+ m_iw_{i-1}\nonumber\\
 &+ v_{i+1}m_i+ v_{i-1}m_i+ w_{i+1}m_i+ w_{i-1}m_i\nonumber\\
 &- m_{i+1}v_i- m_{i-1}v_i- m_{i+1}w_i- m_{i-1}w_i\nonumber\\
 &- v_im_{i+1}- v_im_{i-1}- w_im_{i+1}- w_im_{i-1}\Bigg].
\end{align}

Combining terms, and noting that $v_i=1-w_i-m_i$, we have the dynamics for $m_i$:

\begin{align}
        \frac{dm_i}{dt} &=(P_{MV}^M+P_{VM}^M)m_iv_i\nonumber\\
        &+(P_{MW}^M+P_{WM}^M)m_iw_i\nonumber\\
        &+\frac{D}{(\Delta x)^2}(m_{i+1}-2m_i+m_{i-1}).
\end{align}

Taking the continuum limit in space ($\Delta x \rightarrow 0$), our dynamical equation for $m_i$ becomes a dynamical equation for $m(x,t)$ which we just write as $m$:

\begin{align}\label{eq:dynam_m}
        \frac{d m}{dt} 
        &=(P_{MV}^M+P_{VM}^M)mv\nonumber\\
        &+(P_{MW}^M+P_{WM}^M)mw+D\frac{d^2 m}{dx^2},
\end{align}

where the second derivative in $x$ (\emph{i.e.} the diffusion term) comes from the second-order central differences formula.

A possible extension of the deterministic dynamics is the Chemical Langevin Equation (CLE), which allows for stochastic fluctuations\cite{gillespie2000chemical}. The CLE is a great tool for studying reaction-diffusion systems, and agrees well with simulation methods such as the Gillespie algorithm in many cases. For our system, the equation is given by the following:

\begin{equation}\label{eq:reac_diff}
    \begin{aligned}
        \frac{dx_i}{d t} &= \mathlarger{\sum}\limits_{A\in \mathcal{I}}\mathlarger{\sum}\limits_{B\in \mathcal{I}}\Big[\nu_{AB}^{\mathcal{I}_i}(R)\rho_{AB}^{\mathcal{I}_i}(R)\\
        &+\nu_{AB}^{\mathcal{I}_i}(R)\sqrt{\rho_{AB}^{\mathcal{I}_i}(R)}\eta(0,1)\Big],
    \end{aligned}
\end{equation}

where $\eta(0,1)$ is a Gaussian distributed random variable with mean $\mu=0$ and variance $\sigma^2=1$. 
In terms of the CLE, the asymmetry of payoff tensor components is clearer.
In the ODE (Eq. \ref{eq:reac_diff}), there are infinitely many ways to get a specific value for $P_{IJ}^M+P_{JI}^M$, so if one were to attempt to determine the parameters from data, they may encounter difficulties.

\section{ODE for Survival Probability}

We present a heuristic argument similar to the argument by Lehe \textit{et al.}~\cite{lehe2012rate} to derive the survival probability ($u(x)$) of a new mutation arising in a population of wild-types and vacancies. 

\subsection{\label{app:subsec1} Survival ODE}

In our system, we consider a single mutant that is initiated at position $\xi$ and at time $\tau$. The conditional probability that this mutant is at a position $x$ at a later time $t$ is given by $p(x,t|\xi,\tau)$. In order to simplify calculations we assume that $t\approx\tau$ and that the initial concentrations of wild-types and vacancies are given by their expectation values $\langle w\rangle(x,t)=\langle w\rangle_{init}(x)$ and $\langle v\rangle(x,t)=1-\langle w\rangle_{init}(x,t)=\langle v\rangle_{init}(x)$ respectively. We may decompose the survival probability as follows:

\begin{equation}
    \begin{aligned}
        u(\xi) = \int\limits_{-\infty}^{\infty}u(x)p(x,t|\xi,\tau)dx.
    \end{aligned}
\end{equation}

Taking the derivative with respect to time, we get:

\begin{equation}
    \begin{aligned}
        0 = \int\limits_{-\infty}^{\infty}u(x)\frac{d p(x,t|\xi,\tau)}{d t}dx.
    \end{aligned}
\end{equation}

We replace $p(x,t|\xi,\tau)$ with $m(x,t)$, since the fractional proportion of mutants is the probability that a mutant exists at a location $x$ at time $t$.

\begin{equation}
    \begin{aligned}
        0 &= \int\limits_{-\infty}^{\infty}u\frac{d m}{d t}dx\\
        &=\int\limits_{-\infty}^{\infty}u\Bigg[(P_{MV}^M+P_{VM}^M)mv\\
        &+(P_{MW}^M+P_{WM}^M)mw+D\frac{d^2 m}{d x^2}\Bigg]dx
    \end{aligned}.
\end{equation}

Since we are assuming the short time limit, we write $w(x,t)$ and $v(x,t)$ as $\langle w \rangle_{init}(x)$ and $\langle v \rangle_{init}(x)$ respectively, so we have:

\begin{equation}
    \begin{aligned}
        0 =\int\limits_{-\infty}^{\infty}u\Bigg[&(P_{MV}^M+P_{VM}^M)m\langle v\rangle_{init}\\
        &+(P_{MW}^M+P_{WM}^M)m\langle w\rangle_{init}+D\frac{d^2 m}{d x^2}\Bigg]dx
    \end{aligned}.
\end{equation}

We combine game interaction terms into an expression $\beta(x)$ to simplify the calculations:

\begin{equation}
    \begin{aligned}
        0 &=\int\limits_{-\infty}^{\infty}u\Bigg[\beta m+D\frac{d^2 m}{d x^2}\Bigg]dx.
    \end{aligned}
\end{equation}

Using integration by parts, we get:

\begin{equation}
    \begin{aligned}
        0 &=\int\limits_{-\infty}^{\infty}m\Bigg[\beta u-D\frac{d^2 u}{d x^2}\Bigg]dx\\
        &+D\frac{d}{dx}(um)\Bigg|_{-\infty}^\infty-2D\int\limits_{-\infty}^{\infty}\frac{d m}{d x}\frac{d u}{d x}dx\\
        &=\int\limits_{-\infty}^{\infty}m\Bigg[\beta u-D\frac{d^2 u}{d x^2}\Bigg]dx-2D\int\limits_{-\infty}^{\infty}\frac{d m}{d x}\frac{d u}{d x}dx
    \end{aligned},
\end{equation}

since $\frac{du}{dx}$ and $\frac{dm}{dx}$ are zero at the boundaries. We can additionally make the last term equal to zero if we assume that the mutant distribution is symmetric for small times, and that the survival probability is approximately linear around the nonzero region of the mutant distribution at small times. We thus have:

\begin{equation}
    \begin{aligned}
        0 =\int\limits_{-\infty}^{\infty}m\Bigg[\beta u-D\frac{d^2 u}{d x^2}\Bigg]dx.
    \end{aligned}
\end{equation}

Since $m(x,t)$ may be an arbitrary function, we have:

\begin{equation}
    \begin{aligned}
        0 =\beta u-D\frac{d^2 u}{d x^2}.
    \end{aligned}
\end{equation}

We are interested in the survival probability of a new mutant arising at a location relative to the traveling wild-type wave front. We therefore change to a moving reference frame, traveling at the speed ($s_{w,front}$) of the initial wild-type wave. We thus add a drift term as follows:

\begin{equation}\label{eq:surf1}
    \begin{aligned}
        0 =\beta u-s_{w,front}\frac{d u}{d x}-D\frac{d^2 u}{d x^2}.
    \end{aligned}
\end{equation}

Due to the short time approximation, Eq. (\ref{eq:surf1}) ignores the small probability of two or more mutants establishing in the population. We thus require an additional term in the ODE for the survival probability:

\begin{equation}\label{eq:surf2}
    \begin{aligned}
        \beta u-s_{w,front}\frac{d u}{d x}-D\frac{d^2 u}{d x^2}-F(x,u)=0.
    \end{aligned}
\end{equation}

We will solve for this additional term in \ref{app:subsec3}. Luckily, If one wished to have spatially dependent payoff tensor terms, a simple substitution of $P_{IJ}^K(x)$ can be made in the definition of $\beta(x)$. This may be useful if one wishes to add in a (constantly replenishing) resource, toxin, drug, etc. This is a good approximation when the uptake rate of the spatially varying substance is very slow compared to the velocity of the moving reference frame.

\subsection{\label{app:subsec2} Boundary Conditions}

To solve the survival ODE, we require the boundary conditions $u(\pm\infty)$. At the boundaries, we may use the theory of branching processes. Our state of the branching process is the total number of mutations ($M_{tot}$). The key to determining the boundary conditions are the forward and backward transition rates for the total number of mutants in the system. These rates depend on the sign of the payoff tensor components. If a payoff tensor component is positive, the interaction that it describes contributes to the forward transition rate. Likewise, if the payoff tensor component is negative, then it contributes to the backward transition rate. In addition, the rates in each deme will depend on the number of interacting agents. We sum over the contributions from each deme to obtain the total transition rates for the whole system.
\begin{center}
    \begin{figure*}[!ht]
        \begin{tikzpicture}
            \Vertex[size=1.0,RGB,color={190,174,212},label={$0$}]{A} 
            \Vertex[size=1.0,x=3,RGB,color={190,174,212},label={$1$}]{B} 
            \Vertex[size=1.0,x=6,RGB,color={190,174,212},style={dashed}]{C} 
            \Vertex[size=1.0,x=9,RGB,color={190,174,212},style={dashed}]{D} 
            \Vertex[size=1.0,x=12,RGB,color={190,174,212},label={$LC-1$}]{E} 
            \Vertex[size=1.0,x=15,RGB,color={190,174,212},label={$LC$}]{F}
            % \Edge[Direct,bend=45,label={$r_f$},fontsize={\large}](A)(B)
            \Edge[Direct,bend=45,label={$r_b$},fontsize={\large}](B)(A)
            
            \Edge[Direct,bend=45,label={$r_f$},fontsize={\large},style={dashed}](B)(C)
            \Edge[Direct,bend=45,label={$r_b$},fontsize={\large},style={dashed}](C)(B)
            \Edge[Direct,bend=45,label={$r_f$},fontsize={\large},style={dashed}](C)(D)
            \Edge[Direct,bend=45,label={$r_b$},fontsize={\large},style={dashed}](D)(C)
            \Edge[Direct,bend=45,label={$r_f$},fontsize={\large},style={dashed}](D)(E)
            \Edge[Direct,bend=45,label={$r_b$},fontsize={\large},style={dashed}](E)(D)
            
            \Edge[Direct,bend=45,label={$r_f$},fontsize={\large}](E)(F)
            % \Edge[Direct,bend=45,label={$r_b$},fontsize={\large}](F)(E)
        \end{tikzpicture}            
        \captionsetup{justification=raggedright}
        \caption{A branching process depicting the total number of mutants ($M_{tot}$) with forward transition rates ($r_f$) and backward transition rates ($r_b$)). Here, $L$ is the length of the system, and $C$ is the carrying capacity for a single deme.}
        \label{fig:branching-process}
    \end{figure*}
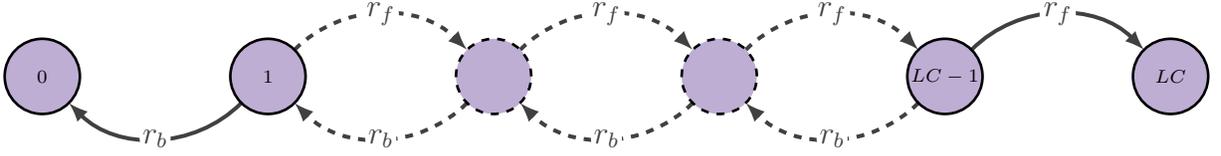
\end{center}
In Fig. \ref{fig:branching-process}, we define $r_b$ as the transition rate from $M_{tot}$ to $M_{tot}-1$, and $r_f$ is the transition rate from $M_{tot}$ to $M_{tot}+1$. We return to the discrete deme picture to derive $r_f$ and $r_b$. Explicitly, these are written in the following way:

\begin{equation}
    \begin{aligned}
        r_b = \mathlarger{\sum}\limits_{i=1}^L &[\phi^-(P_{MV}^M)
        + \phi^-(P_{VM}^M)]m_i\langle v\rangle_{init,i} \\
        &+ [\phi^-(P_{MW}^M)+ \phi^-(P_{WM}^M)]m_i\langle w\rangle_{init,i} \\
        r_f = \mathlarger{\sum}\limits_{i=1}^L &[\phi^+(P_{MV}^M)+ \phi^+(P_{VM}^M)]m_i\langle v\rangle_{init,i}  \\
        &+ [\phi^+(P_{MW}^M)+ \phi^+(P_{WM}^M)]m_i\langle w\rangle_{init,i}, 
    \end{aligned}
\end{equation}

where we make use of the ``ReLU'' function (denoted by $\phi^+(x)$). Alone, this function picks out the value $x$ if $x\ge0$, and returns $0$ otherwise. When written as $\phi^-(x)$, the value $-x$ is returned if $x\le0$. Thus, if a payoff tensor component is negative, it contributes to the backward transition rate, and if it is positive, it contributes to the forward transition rate. 

To find the values of the transition rates at the boundaries, we shift our spatial region where we are either far into the bulk of $\langle w\rangle_{init} $ or far ahead of the bulk of $\langle w\rangle_{init} $ (the bulk of $\langle v\rangle_{init} $). Far in the bulk of the wild-type wave, we have the following:

\begin{equation}
    \begin{aligned}
        r_b^{(-\infty)} = \mathlarger{\sum}\limits_{i=1}^L &[\phi^-(P_{MW}^M) + \phi^-(P_{WM}^M)]m_i\langle w\rangle_{init,i} \\
        r_f^{(-\infty)} = \mathlarger{\sum}\limits_{i=1}^L &[\phi^+(P_{MW}^M)+ \phi^+(P_{WM}^M)]m_i\langle w\rangle_{init,i}.
    \end{aligned}
\end{equation}

We define the quantity $\gamma$, which is the ratio of the backward to forward transition rates ($\gamma=r_b/r_f$). Far in the bulk of the wild-type wave, we have:
\begin{equation}
    \begin{aligned}
    &\gamma^{(-\infty)}=\frac{r_b^{(-\infty)} }{r_f^{(-\infty)} }\\
    &=\frac{\mathlarger{\sum}\limits_{i=1}^L [\phi^-(P_{MW}^M) + \phi^-(P_{WM}^M) ]m_i\langle w\rangle_{init,i}}{\mathlarger{\sum}\limits_{i=1}^L [\phi^+(P_{MW}^M)+ \phi^+(P_{WM}^M)]m_i\langle w\rangle_{init,i} }\\
    &=\frac{\phi^-(P_{MW}^M)+ \phi^-(P_{WM}^M)}{\phi^+(P_{MW}^M)+ \phi^+(P_{WM}^M)}.
    \end{aligned}
\end{equation}
Similarly, far ahead of the bulk of the wild-type wave, we have:
\begin{equation}
    \begin{aligned}
    \gamma^{(\infty)}=\frac{\phi^-(P_{MV}^M)+ \phi^-(P_{VM}^M)}{\phi^+(P_{MV}^M)+ \phi^+(P_{VM}^M)}.
    \end{aligned}
\end{equation}
The survival probability is then given by the well known expression for the survival probability, beginning with a single individual \cite{nowak2006evolutionary}:
\begin{align}\label{eq:BCS_eq}
        u(\pm\infty)&=\frac{1}{1-\sum\limits_{j=1}^{LC-1}\prod\limits_{k=1}^{j}\gamma^{(\pm\infty)}}\nonumber\\
        &=\frac{1}{1-\sum\limits_{j=1}^{LC-1}(\gamma^{(\pm\infty)})^j}\nonumber\\
        &=\frac{1-\gamma^{(\pm\infty)}}{1-(\gamma^{(\pm\infty)})^{LC}}\nonumber\\
        &\approx\lim\limits_{L\rightarrow\infty}\frac{1-\gamma^{(\pm\infty)}}{1-(\gamma^{(\pm\infty)})^{LC}}.
\end{align}
\subsection{\label{app:subsec3} Finding a form for $F(x,u)$}

We may now make an argument for the form of $F(x,u)$ in Eq. (\ref{eq:surf2}). We assume that the values of the survival probability saturate to constants in regions near the boundaries of the system. In this case, we have $\frac{d u}{d x}=0$ and $\frac{d^2 u}{d x^2}=0$. Denoting $\tilde{x}$ as either $+\infty$ or $-\infty$ and $\tilde{u}=u(\tilde{x})$, we have:
\begin{equation}\label{eq:surf5}
    \begin{aligned}
        \beta(\tilde{x})\tilde{u}-F(\tilde{x},\tilde{u})=0.
    \end{aligned}
\end{equation}
Furthermore, at the boundaries, either $\langle v \rangle_{init}(-\infty)=0$ and $\langle w \rangle_{init}(-\infty)=1$ or $\langle v \rangle_{init}(\infty)=1$ and  $\langle w \rangle_{init}(\infty)=0$. We may thus write a general expression:
\begin{equation}\label{eq:surf6}
    \begin{aligned}
        (a+b)\tilde{u}-F(\tilde{x},\tilde{u})=0.
    \end{aligned}
\end{equation}
Here, $a$ and $b$ replace the payoff tensor components. 
We can also write specific values for the boundary conditions depending on the signs of the payoff tensor components. 
Due to the similar form for $\gamma^{(\infty)}$ and $\gamma^{(-\infty)}$, we may write general expressions. More explicitly, we write:
\begin{equation}
    \begin{aligned}
    \tilde{\gamma}=\frac{\phi^-(a) + \phi^-(b)}{\phi^+(a)+ \phi^+(b)}
    \end{aligned}
\end{equation}
and
\begin{equation}
    \begin{aligned}
        \tilde{u}=\frac{1-\tilde{\gamma}}{1-\tilde{\gamma}^{LC}}.
    \end{aligned}
\end{equation}
We then proceed, looking at specific relationships between $a$ and $b$, with the results summarized in Tab. (\ref{tab:phase_surf_table})
\begingroup
\renewcommand{\arraystretch}{10} % Default value: 1
    % \begin{widetext}
       \begin{center}
        \begin{table*}[ht]\label{eq:surftable}
            \centering
            \begin{tblr}{ |c|c|c|c|c|c|c| }
            \hline
            $\textrm{sgn}(a)$ & $\textrm{sgn}(b)$ & $\textrm{sgn}(1-|b|/|a|)$ & $\tilde{\gamma}$ & $\tilde{u}$ & $\lim\limits_{L \to \infty}\tilde{u}$ & $\lim\limits_{L \to \infty}F(\tilde{x},\tilde{u})$\\
            \hline
            $+$ & $+$ & $\forall$ & $0$ & $1$ & $1$ & $\begin{aligned}
                &a+b\\
                =(&a+b)\tilde{u}^2
            \end{aligned}$\\
            \hline[dashed]
            $+$ & $-$ & $-$ & $-\frac{b}{a}$ & \mbox{\Large\(\frac{1+\frac{b}{a}}{1-(-\frac{b}{a})^{LC}}\)} & $0$ & $0=0\tilde{u}^2$\\
            \hline[dashed]
            $+$ & $-$ & $+$ & $-\frac{b}{a}$ & \mbox{\Large\(\frac{1+\frac{b}{a}}{1-(-\frac{b}{a})^{LC}}\)} & $1+\frac{b}{a}$ & $\begin{aligned}
                (&a+b)\Big(1+\frac{b}{a}\Big)\\
                =&a\Big(1+\frac{b}{a}\Big)\Big(1+\frac{b}{a}\Big)\\
                =&a\tilde{u}^2
            \end{aligned}$\\
            \hline[dashed]
            $-$ & $-$ & $\forall$ & $\infty$ & $0$ & $0$ & $0=0\tilde{u}^2$\\
            \hline[dashed]
            $-$ & $+$ & $+$ & $-\frac{a}{b}$ & \mbox{\Large\(\frac{1+\frac{a}{b}}{1-(-\frac{a}{b})^{LC}}\)} & $0$ & $0=0\tilde{u}^2$\\
            \hline[dashed]
            $-$ & $+$ & $-$ & $-\frac{a}{b}$ & \mbox{\Large\(\frac{1+\frac{a}{b}}{1-(-\frac{a}{b})^{LC}}\)} & $1+\frac{a}{b}$ & $\begin{aligned}
                (&a+b)\Big(1+\frac{a}{b}\Big)\\
                =&b\Big(1+\frac{a}{b}\Big)\Big(1+\frac{a}{b}\Big)\\
                =&b\tilde{u}^2
            \end{aligned}$\\
            \hline
            \end{tblr}
        \caption{The phenomenological correction term across the phase space of asymmetric game interactions.}
        \label{tab:phase_surf_table}
        \end{table*}       
    \end{center} 
    % \end{widetext}   
\endgroup

% \clearpage
% \newpage
% \mbox{~}
% \clearpage
% \vspace{10in}
We may thus write $F(\tilde{x},\tilde{u})$ at one of the boundaries:

\begin{equation}
    \begin{aligned}
        F(\tilde{x},\tilde{u}) = (\phi^+(a)+\phi^+(b))\tilde{u}^2.
    \end{aligned}
\end{equation}

To interpolate between the left and right boundaries, we make the Ansatz:

\begin{equation}\label{eq:F-function}
    \begin{aligned}
        F(x,u) &= [(\phi^+(P_{MV}^M)+\phi^+(P_{VM}^M))\langle v \rangle_{init}(x)\\
        &+(\phi^+(P_{MW}^M)+\phi^+(P_{WM}^M))\langle w \rangle_{init}(x)]u(x)^2.
    \end{aligned}
\end{equation}

Plugging Eq. (\ref{eq:F-function}) into Eq. (\ref{eq:surf2}) and without writing the $x$ dependence for compactness, we have:

\begin{equation}\label{eq:surf7}
    \begin{aligned}
        &[(P_{MV}^M+P_{VM}^M)\langle v\rangle_{init}\\
        &+(P_{MW}^M+P_{WM}^M)\langle w\rangle_{init}]u-s_{w,front}\frac{d u}{d x}-D\frac{d^2 u}{d x^2}\\
        &-[(\phi^+(P_{MV}^M)+\phi^+(P_{VM}^M))\langle v \rangle_{init}\\
        &+(\phi^+(P_{MW}^M)+\phi^+(P_{WM}^M))\langle w \rangle_{init}]u^2=0.
    \end{aligned}
\end{equation}

If one wished to reduce the game interactions to symmetric game interactions as apposed to asymmetric game interactions, one would simply set $P_{WM}^M=P_{MW}^M$ and $P_{VM}^M=P_{MV}^M$ to give:

\begin{equation}\label{eq:surf5_sym}
    \begin{aligned}
        &2[P_{MV}^M\langle v\rangle_{init}
        +P_{MW}^M\langle w\rangle_{init}]u-s_{w,front}\frac{d u}{d x}-D\frac{d^2 u}{d x^2}\\
        &-2[\phi^+(P_{MV}^M)\langle v \rangle_{init}+\phi^+(P_{MW}^M)\langle w \rangle_{init}]u^2=0.
    \end{aligned}
\end{equation}

A simple redefinition of the payoff tensor components can be done by absorbing the factor of $2$ as $2P_{IJ}^K=\pi_{IJ}^K$ to give:

\begin{equation}\label{eq:surf6_sym}
    \begin{aligned}
        &[\pi_{MV}^M\langle v\rangle_{init}+\pi_{MW}^M\langle w\rangle_{init}]u
        -s_{w,front}\frac{d u}{d x}-D\frac{d^2 u}{d x^2}\\
        &-[\phi^+(\pi_{MV}^M)\langle v \rangle_{init}+\phi^+(\pi_{MW}^M)\langle w \rangle_{init}]u^2=0.
    \end{aligned}
\end{equation}

Based on the boundary condition equations (Eq. \ref{eq:BCS_eq}), our $\gamma$'s are either $0$ or $\infty$, making our boundary conditions equal to either $1$ or $0$ respectively. 
If we include diagonal terms in our payoff tensor, as in the following equation, we may again observe the full range of boundary conditions from $0$ to $1$.
The symmetric form of Eq. (\ref{eq:surf7}) with diagonal payoff tensor terms is as follows:

\begin{equation}\label{eq:surf6_sym_self_interactions}
    \begin{aligned}
        &[\pi_{MV}^M\langle v\rangle_{init}+\pi_{MW}^M\langle w\rangle_{init}+\pi_{MM}^M\langle m\rangle_{init}]u\\
        &-s_{w,front}\frac{d u}{d x}-D\frac{d^2 u}{d x^2}-[\phi^+(\pi_{MV}^M)\langle v \rangle_{init}\\
        &+\phi^+(\pi_{MW}^M)\langle w \rangle_{init}
        +\phi^+(\pi_{MM}^M)\langle m \rangle_{init}]u^2=0.
    \end{aligned}
\end{equation}

If one wished to reduce Eq. (\ref{eq:surf7}) to have neutral interactions between mutants and wild-types (the dashed line in Fig. \ref{fig:Clock_Plot}), one would set $P_{WM}^M = -P_{MW}^M$ and $P_{WM}^W = -P_{MW}^W$, obtaining:

\begin{equation}\label{eq:surf7_nogame}
    \begin{aligned}
        &[(P_{MV}^M+P_{VM}^M)\langle v\rangle_{init}]u-s_{w,front}\frac{d u}{d x}-D\frac{d^2 u}{d x^2}\\
        &-[(\phi^+(P_{MV}^M)+\phi^+(P_{VM}^M))\langle v \rangle_{init}\\
        &+(\phi^+(P_{MW}^M)+\phi^+(P_{WM}^M))\langle w \rangle_{init}]u^2=0.
    \end{aligned}
\end{equation}

Plugging in the values $P_{MV}^M=1,P_{VM}^M=-(1-r_m),P_{MW}^M=1,$ and $P_{WM}^M=-1$, we obtain the survival probability ODE from \cite{lehe2012rate}:

\begin{equation}\label{eq:surf8_nogame}
    \begin{aligned}
        r_m\langle v\rangle_{init}u
        -s_{w,front}\frac{d u}{d x}-D\frac{d^2 u}{d x^2}
        -u^2=0
    \end{aligned}.
\end{equation}

For a system that is truly without game interactions between mutants and wild-types, one would set $P_{WM}^M = P_{MW}^M = P_{WM}^W = P_{MW}^W = 0$, to obtain:

\begin{equation}\label{eq:surf9_truenogame}
    \begin{aligned}
        &[(P_{MV}^M+P_{VM}^M)\langle v\rangle_{init}]u
        -s_{w,front}\frac{d u}{d x}-D\frac{d^2 u}{d x^2}\\
        &-(\phi^+(P_{MV}^M)+\phi^+(P_{VM}^M))\langle v \rangle_{init}u^2=0.
    \end{aligned}
\end{equation}

\end{document}